\documentclass[twocolumn]{aastex6}

\usepackage{graphicx}
\usepackage{hyperref}
\usepackage{amssymb,amsmath}
\usepackage{mathtools}
\usepackage{stmaryrd}

\def\lea{\mathrel{<\kern-1.0em\lower0.9ex\hbox{$\sim$}}}
\def\gea{\mathrel{>\kern-1.0em\lower0.9ex\hbox{$\sim$}}}
\newcommand{\lta}{{\>\rlap{\raise2pt\hbox{$<$}}\lower3pt\hbox{$\sim$}\>}}
\newcommand{\gta}{{\>\rlap{\raise2pt\hbox{$>$}}\lower3pt\hbox{$\sim$}\>}}

\begin{document}
\title{Chandra X-ray Observatory and Hubble Space Telescope observations of the intermediate-age cluster GLIMPSE$-$C01}

\author{Jeremy Hare,\altaffilmark{1,2} Oleg Kargaltsev,\altaffilmark{1,2} Blagoy Rangelov\altaffilmark{3}}

\altaffiltext{1}{Department of Physics, The George Washington University, 725 21st St, NW, Washington, DC 20052}
\altaffiltext{2}{The George Washington Astronomy, Physics, and Statistics Institute of Sciences (APSIS)}
\altaffiltext{3}{{ Texas State University, Department of Physics, 601 University Drive, San Marcos, TX 78666}}
\email{jeh86@gwu.edu}

\shorttitle{GLIMPSE$-$C01}

\begin{abstract}
We report the results of {\sl Hubble Space Telescope (HST)}  and {\sl Chandra X-ray Observatory (CXO)} observations of the { 	GLIMPSE-C01 (hereafter GC01) star cluster}. Color-magnitude and color-color diagrams  suggest a cluster age of { $\gtrsim2$ Gyrs up to $\sim10$ Gyrs (dependent on GC01's metallicity), a distance of 3.3-3.5 kpc, and strong differential reddening with $A_V=14-22$.}   After performing astrometric corrections, we find that nine of the 15 X-ray sources have at least one NIR counterpart within the { 2$\sigma$} {\sl CXO} positional error circles. { However, given the very high density of NIR sources in the cluster, most of these counterparts are likely due to chance coincidence.  We jointly analyze the X-ray and NIR properties to assess the likelihood of true associations. Based primarily on their X-ray properties, we identify} an LMXB candidate (source X2), a CV candidate (source X1), and { an AB candidate}  (source X9). Source X11 is detected during an X-ray flaring episode with a flare luminosity ($L_{X}=2.1\times10^{33}$ erg s$^{-1}$) and has a quiescent luminosity $L_{X}<8.0\times10^{30}$ erg s$^{-1}$, in 0.5--8 keV at the distance of  GC01, suggesting that the source is { either} an { AB or CV}. We also discuss the limits on an intermediate mass black hole at the center of GC01 and the challenges of X-ray source classification imposed by the limitations of the existing data and instrumentation along with future prospects in the {\sl James Webb Space Telescope} era.
\end{abstract}


\section{Introduction}

Thousands of star clusters exist within the Milky Way Galaxy and galactic halo \citep{2015A&A...581A..39S}. Typically these clusters are classified either as old globular clusters (GCs), which have high concentrations of stars, or open clusters, { which are less dense}. Open clusters, typically found in the galactic disk,  have masses on the order of 10$^{3}$ M$_{\odot}$, { near-solar metallicities}, and ages ranging between 1 Myr and 1 Gyr, although several old open clusters have also been discovered (e.g., NGC 6791, NGC 188; \citeauthor{2005A&A...438.1163K} \citeyear{2005A&A...438.1163K}). Among these are the Young Massive Clusters (YMCs), which are sometimes considered to be a class on their own, with masses $\gtrsim 10^4 M_{\odot}$ and ages up to a few hundred Myrs \citep{2010ARA&A..48..431P}. Low-metallicity GCs are mostly found in  the Galactic halo (although higher metallicity GCs exist in the disk) with masses $\sim10^{4}$--$10^{6}$ M$_{\odot}$ and  ages of $8-14$ Gyrs (see e.g., Fig~9 in \citealt{2010ApJ...708..698D}).

The GLIMPSE$-$C01 cluster  (GC01, hereafter), which was discovered with \emph{Spitzer} during the Galactic Legacy Infrared Mid Plane Survey \citep{2005AJ....129..239K}, is interesting because it does not fit well into any of these categories. GC01 is located at $l=31.3^{\circ}$,  $b=-0.1^{\circ}$, and lies  within 10~pc of the Galactic mid-plane. It was originally estimated to have a mass of $\sim 10^{5}$ M$_{\odot}$, a half-light radius of 36$''$, and a distance of 3.1-5.2 kpc \citep{2005AJ....129..239K}. This cluster is also highly reddened  with  $A_V\sim15\pm3$, which likely varies across the cluster (\citeauthor{2005AJ....129..239K} \citeyear{2005AJ....129..239K}; \citeauthor{2005A&A...442..195I} \citeyear{2005A&A...442..195I}; \citeauthor{2016AJ....152..173D} \citeyear{2016AJ....152..173D}).

Since the discovery of GC01, widely varying  distance and age estimates have been reported. Both \cite{2005A&A...442..195I} and \cite{2016AJ....152..173D} used red clump stars, observed in slightly different filters, to estimate a distance to GC01 of 3.7 and 5.2 kpc, respectively.  \cite{2011MNRAS.411.1386D} used near infrared spectroscopy of 50 stars in the cluster to calculate their velocities and, by assuming that GC01 is moving with the disk, derive a kinematic distance of 5.0$\pm0.9$ kpc. The measured velocity dispersion of the stars suggests a virial mass of (8$\pm3)\times$10$^{4}$ M$_{\odot}$ \citep{2011MNRAS.411.1386D}. 

Due to its centrally-concentrated appearance in the NIR images, GC01 strongly resembles GCs.  Therefore,  it was initially suggested that GC01 is a GC with an age of at least a few Gyr. The diffuse infrared emission coincident with GC01 in both {\sl Spitzer} IRAC and MIPS images was interpreted in support of a GC passing  through the Galactic disk and interacting with the gas and dust in the disk \citep{2005AJ....129..239K}. The advanced age and classification as a globular cluster would be consistent with the lack of radio emission (typically seen in younger open clusters), its high central stellar density, and the large number of giant branch stars with no luminous supergiants \citep{2005AJ....129..239K}. However, \cite{2011MNRAS.411.1386D} found that GC01 is more compact than typical GCs of a similar mass and that the mass density is more similar to that of YMCs.  Using the K-band mass-to-light ratio \cite{2011MNRAS.411.1386D} infer an age between 0.3 and 2 Gyr. Models of the Red Giant Branch (RGB) tip brightness in the color-magnitude diagram (CMD) for stars in GC01 also suggests an age between 1-2.5 Gyr \citep{2016AJ....152..173D}. Therefore, GC01 could be a rare intermediate age massive cluster \citep{2011MNRAS.411.1386D}. 
  
  The origin of massive GCs is a matter of ongoing debate \citep{2017MNRAS.465.3622R}. It has been suggested that at least some  massive GCs in the disk could be the outcome of YMCs evolution \citep{2010ARA&A..48..431P}.  However, GCs could have also coalesced out of a primordial gas cloud that later collapsed into the Galactic disc (see e.g. \citealt{2003Sci...299...65K}). Furthermore, recent simulations show that low-metallicity GCs may represent the cores of satellite galaxies that merged with the Milky Way \citep{2017MNRAS.465.3622R}.  If the age of GC01 is indeed substantially less than the age of the Galaxy, its metallicity is closer to solar \citep{2013MNRAS.436..122L}, and its X-ray binary population is different from those in GCs, then it may represent a missing evolutionary link  between the  YMCs and massive GCs. { On the other hand, if GC01 has a low metallicity and old age (as suggested by \citealt{2005A&A...442..195I} and \citealt{2005AJ....129..239K}, respectively), then it could be similar to the rare GCs that reside in the Milky Way disk outside of the bulge, such as NGC 6544 \citep{2014AJ....148...18C} and Glimpse-C02 \citep{2008A&A...489..583K}.}

 Studies of X-ray sources located in open and globular clusters are crucial for understanding the evolution, dynamics, and stellar populations of these objects (\citeauthor{2013ASPC..470..251V} \citeyear{2013ASPC..470..251V}; \citeauthor{2010PNAS..107.7164P} \citeyear{2010PNAS..107.7164P}, \citeauthor{2010AIPC.1314..135H} \citeyear{2010AIPC.1314..135H}). Typically, the X-ray populations of globular and aged open clusters consist of cataclysmic variables (CVs), non-accreting\footnote{These MPSs can be solitary, in wide binaries, or in binaries with a very low mass companion which is ablated by the pulsar's wind (i.e., redbacks or blackwidows; for a recent review see \citealt{2017JApA...38...42M})} millisecond radio pulsars (MSPs), neutron star (NS) or black hole (BH) low mass X-ray binaries which can be in quiescence (LMXBs or qLMXBs, respectively), and active binaries (ABs, such as RS CVn and W Uma type systems; \citeauthor{2013ASPC..470..251V} \citeyear{2013ASPC..470..251V}; \citeauthor{2010PNAS..107.7164P} \citeyear{2010PNAS..107.7164P}, \citeauthor{2010AIPC.1314..135H} \citeyear{2010AIPC.1314..135H}).  In dense GCs the relative numbers of  these  objects are expected  to be related to the number of dynamical encounters  (see \citeauthor{2006ApJ...646L.143P} \citeyear{2006ApJ...646L.143P}, \citeauthor{2006ApJ...651.1098H} \citeyear{2006ApJ...651.1098H}, and references therein). In less-dense clusters the X-ray source population is more likely to be primordial in nature (see e.g., \citeauthor{2006ApJ...647.1065K} \citeyear{2006ApJ...647.1065K}, \citeauthor{2010ApJ...712..380L} \citeyear{2010ApJ...712..380L}).  As a result, GCs tend to be rich with recycled  MSPs, many of which are in binaries (e.g., \citealt{2006ApJ...646.1104B}), while in old ($>7$ Gyr) open clusters the observed dominating X-ray source population is ABs (see e.g., \citealt{2017ApJ...837..130V}). If GC01 is indeed an intermediate age massive cluster, the population of X-ray sources would be essentially unknown. This prompted us to carry out multi-band \emph{Hubble Space Telescope} ({\sl HST}) observations to look for counterparts of the X-ray sources seen in the archival \emph{Chandra X-ray Observatory} ({\sl CXO}) image of  GC01\footnote{The initial analysis of the {\sl CXO} data was reported by  \cite{2007arXiv0708.3365P}. }. Below we report the results of these  {\sl HST} observations together with a re-analysis of the archival {\sl CXO} data and a discussion of the existing limitations and challenges imposed by the current instrumentation. 

We present {\sl HST} and {\sl CXO} observations, and the data
reduction  methods  in  Section \ref{obs}. In Section \ref{analysis} we  describe the analysis of  the {\sl CXO}  and  {\sl HST} data, including the  properties of the X-ray sources, NIR/optical photometry,  and the cross-correlation of optical/NIR counterparts with the X-ray source positions.  We describe  the properties of the cluster  in Section \ref{clusprop}. The multi-wavelength classification of X-ray sources and the limit on the mass of a possible intermediate mass BH (IMBH) are discussed in Section \ref{xray}. We discuss the current limitations of this study in Section \ref{outlook} and summarize the results of our findings in Section \ref{summ}.

\begin{figure*}
\begin{center}
\includegraphics[scale=0.5]{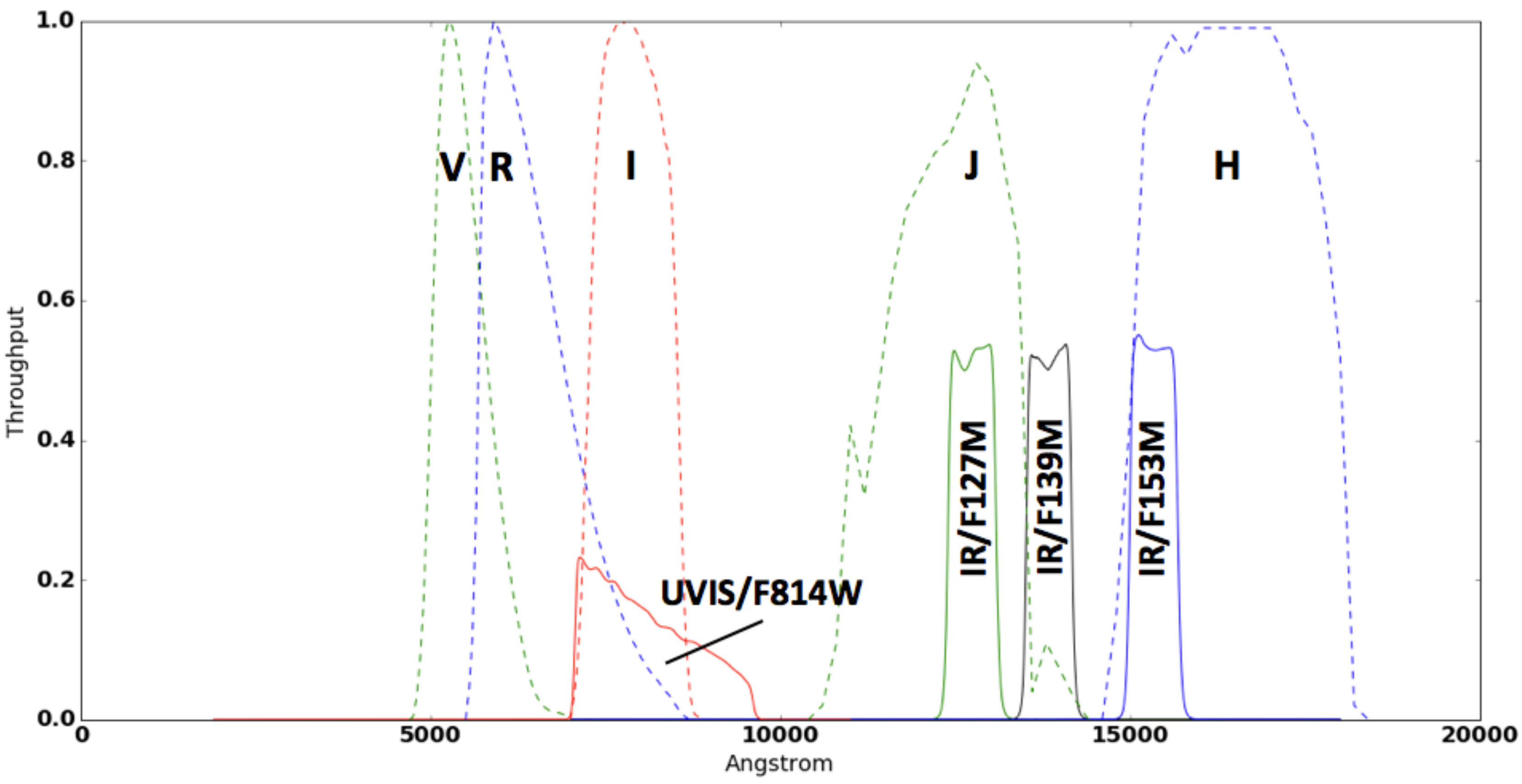}
\caption{Comparison of the {\sl HST} WFC3 IR and UVIS filters used in this study with Johnson filters.}
\label{filter}
\end{center}
\end{figure*}

\begin{figure*}
\includegraphics[scale=0.5,{trim=-0 0 0 0}]{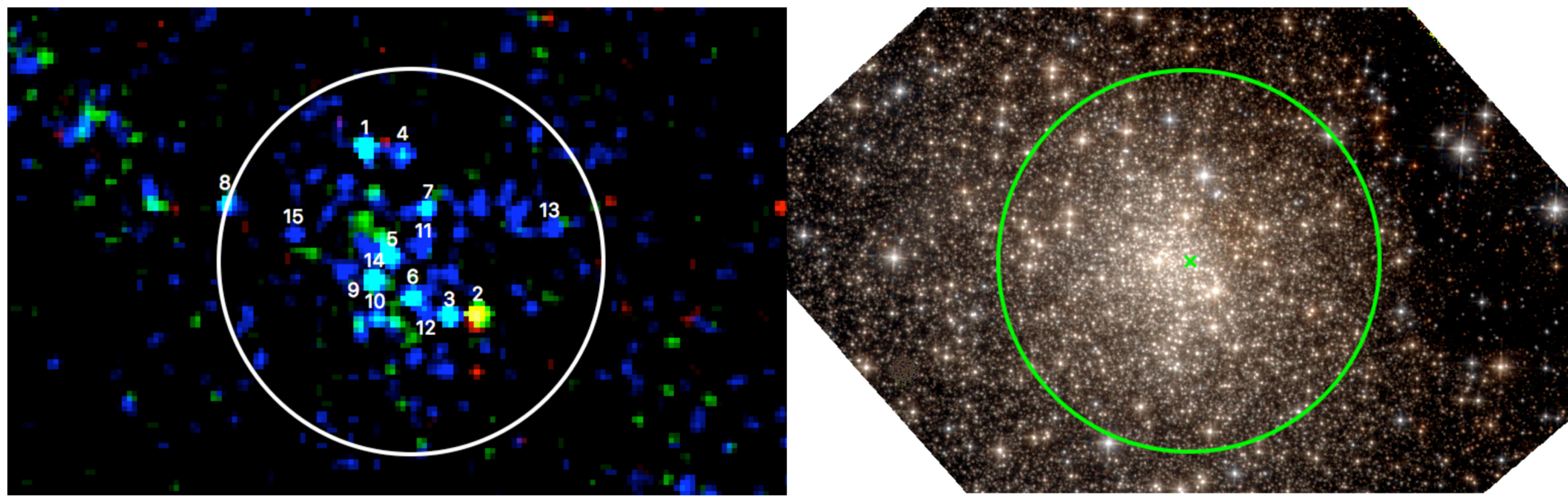}
\caption{ \emph{CXO}  and {\sl HST} images of the GC01 field showing the same area on the sky (North is up, East is to the left). { The solid white (left) and green (right) circles represent the $36''$ half-light radius, while the green cross (right) shows the cluster center \citep{2005AJ....129..239K}.} {\sl Left:} Binned (by a factor of 2) and smoothed (with a Gaussian kernel with a radius of 2$''$) \emph{CXO} false color image  ($3-8$ keV - blue, $1.5-3$ keV -- green, and $0.5-1.5$ keV -- red). X-ray sources detected with a significance $>$6 net counts are numbered in correspondence with Table \ref{xsrc}.  {\sl Right:} False color {\sl HST} image made from the F127M (blue), F139M (green), and F153M (red) WFC3/IR images. The magenta cross marks the cluster center determined by Kobulnicky et al.\ (2005). 
}\label{ROI}
\end{figure*}
\vspace{2mm}

\section{Observations and  Data Reduction}
\label{obs}
\subsection{HST WFC3/UVIS and WFC3/IR Data}

GC01 was observed by \emph{HST} during a single-orbit  visit on 2015 October 5 (GO 14183; see  Table~\ref{hstobs} for details). The program consisted of imaging observations with the Wide Field Camera 3 (WFC3), using both the Ultraviolet-Visible  (UVIS; $162''\times162''$ field of view with a 0\farcs04 pixel scale) and the  Infrared  (IR; $136''\times123''$ field of view with a 0\farcs13 pixel scale) channels. The UVIS F814W (wide $I$) filter and a set of medium NIR filters (F127M, F139M, and F153M) were used (see Figure \ref{filter} for filter throughputs) { and the exposure times for each filter are provided in Table \ref{hstobs}}. The data were downloaded from the Mikulski Archive for Space Telescopes (MAST\footnote{http://archive.stsci.edu/}). The final images are produced by PyRAF's $Multidrizzle$ task which aligns images from separate exposures and performs geometric distortion corrections.

The CCDs of the UVIS detector have degraded overtime due to their exposure to high energy particles in low-Earth orbit. This degradation causes streaking (i.e., leakage of the charge from a source pixel into adjacent pixels) in the direction of the readout resulting in losses in the charge transfer efficiency (CTE). These losses can lead to errors in the photometry and also affects the image quality. MAST provides UVIS images with the empirically-calculated\footnote{See \url{http://www.stsci.edu/hst/wfc3/ins_performance/CTE/} for more details.}  CTE corrections applied. We used these CTE-corrected UVIS images for the photometric  measurements reported below.

The high density of stars in GC01 makes PSF fitting photometry the only robust way to measure the positions and photometric magnitudes of the stars in the cluster. To perform the PSF fitting photometry we used the DOLPHOT\footnote{http://americano.dolphinsim.com/dolphot/} package \citep{2000PASP..112.1383D}. DOLPHOT has a number of adjustable parameters, which can influence the photometric measurements. We chose settings  similar to those listed in Table 4 of \cite{2012ApJS..200...18D} with the exception of the FitSky parameter value. This parameter fixes how the background is determined for the PSF fit. We adopted  FitSky=2 since it is recommended\footnote{\label{note1}\url{http://americano.dolphinsim.com/dolphot/dolphot.pdf}} for crowded fields. We had to apply a number of parameter cuts to the catalog produced by DOLPHOT 
to remove the spurious sources and those with inaccurate photometry (e.g., saturated and crowded sources).  These cuts  were chosen differently for different purposes and the specific choices are described below.

\begin{deluxetable}{crlc}
\tablecaption{\emph{HST} WFC3 exposures\label{hstobs}}
\tablehead{
\colhead{Order} & \colhead{Channel} & \colhead{Filter} & \colhead{Exposure\  (s)}}
\startdata
1 & IR & F127M & 548 \\
2 & IR & F139M & 598 \\
3 & IR & F153M & 598 \\
4 & UVIS & F814W & 540 \\
\enddata
\tablecomments{All exposures were taken within a single {\sl HST} orbit in the order specified in the Table.}
\end{deluxetable}

\subsection{CXO ACIS-S3 Data}
\label{cacis}

We have downloaded and reanalyzed the archival {\sl CXO} observation of GC01. The  data were taken on 15 August 2006 (ObsID 6587) with the Advanced CCD Imaging Spectrometer (ACIS).  The exposure time was 45.9 ks and GC01 was imaged on the back-illuminated S3 chip. The detector was operated in full-frame mode, which
provides a time resolution of 3.24 s, with the `very
faint' telemetry format. We restricted the X-ray data analysis to the standard 0.5--8 keV energy range. To extract the properties of X-ray sources we used the  CIAO tools\footnote{Chandra Interactive Analysis Tools (CIAO) version 4.9} and the {\sl ACIS Extract}\footnote{See \url{http://personal.psu.edu/psb6/ACIS/acis_analysis.html} for details.} package \citep{2010ApJ...714.1582B,2012ascl.soft03001B} for our analysis.

{ The X-ray sources are detected by locating local maxima in the reconstructed {\sl CXO} image, which is created using the Lucy-Richardson image reconstruction algorithm \citep{1974AJ.....79..745L}. The iterative source detection procedure and strategy, which is optimized for finding faint X-ray sources in crowded fields, is outlined in \citep{2010ApJ...714.1582B,2018arXiv180204902T}. While determining the source's position, the location of the source on the chip and its proximity  to other sources is taken into account. Source positions for uncrowded, on-axis sources (all sources considered below are close to being on-axis) are determined by calculating the centroid of the events. If there are two or more sources near one another, finding the peak in a local maximum-likelihood image reconstruction is the most accurate way to recover the source positions \citep{2010ApJ...714.1582B}. This procedure allowed us to accurately recover the positions of sources X9 and X14, which are located only $\sim$0\farcs6 from each other and were reported as a single source (i.e., X7) in \cite{2007arXiv0708.3365P}.}

{\sl ACIS Extract} provides fluxes for all sources, which are calculated by multiplying the median energy of the counts by the number of net counts and dividing by the exposure time and an aperture corrected mean effective area \citep{2010ApJ...714.1582B}. For the six sources that have $>$30 net counts in the 0.5--8.0 keV band, we also extracted the spectra from $r=1''$ circular apertures. The spectra were extracted using the CIAO tool {\tt specextract} with an aperture correction applied to the effective area\footnote{See \url{http://http://cxc.harvard.edu/ciao/threads/pointlike/}}. We then fit an absorbed power-law (PL) model to characterize their spectra.  No background subtraction was performed because of its negligible contribution (i.e., $<{ 5}\%$ of the { total} source counts for each of the sources { in the 0.5-8 keV energy range}). Due to the small number of counts in all fits, we fixed the hydrogen absorption column to $N_{\rm H}=4\times10^{22}$ cm$^{-2}$, which corresponds to an extinction of $A_V=$18 (a plausible value  for GC01; see Section \ref{colcol}) assuming the $N_{\rm H}$--$A_V$ relation from \cite{2009MNRAS.400.2050G}. The X-ray spectra were fit using XSPEC version 12.9.1 \citep{1996ASPC..101...17A}. { Throughout the paper, the photoelectric absorption cross-sections of \cite{1996ApJ...465..487V} were used with solar abundances \citep{1989GeCoA..53..197A} to account for absorption by neutral gas.}

\subsection{Astrometric Corrections}
\label{astrometry}
In order to accurately identify the WFC3/IR and UVIS counterparts to the X-ray sources we corrected the absolute astrometry of both the {\sl HST} and {\sl CXO} images. For the HST images we matched stars from the Two Micron All-Sky Survey catalog (2MASS; \citealt{2006AJ....131.1163S,2012wise.rept....1C}) to stars in the field of view of the WFC3/IR. We chose 2MASS due to the overlap of the 2MASS and WFC3/IR filters. We then selected 24 stars from the F127M image with bright and unblended 2MASS counterparts.  All selected stars are located outside the dense cluster core to avoid source confusion.  Prior to the correction, a systematic offset of $0\farcs37$ ($\Delta\alpha=-0\farcs37, \Delta\delta=0\farcs02$)  was present between the 2MASS and WFC3/IR stars. We then calculated the astrometric offset between the sources and applied the coordinate transformation to the drizzled images. A subset of 11 of the original 24 WFC3/IR sources, which had unsaturated counterparts in the WFC3/UVIS images, were then used to align the WFC3/UVIS image to the WFC3/IR image. 

For {\sl CXO}, we found 10 sources in the 0.5--8 keV band ACIS image with $>$7 net counts that also had 2MASS counterparts and were located outside of the cluster's core. We used the CIAO tools {\tt wcs\_match} to calculate the astrometric solution and {\tt wcs\_update} to apply the coordinate transformation. An offset of $0\farcs36$ ($\Delta\alpha=-0\farcs35, \Delta\delta=0\farcs07$) existed prior to the astrometric correction. { After applying the corrections to both {\sl CXO} and {\sl HST}, we calculated the root-mean square residuals of the alignments and added them in quadrature. The resulting $0\farcs1$ value is adopted as the $1\sigma$ uncertainty of { the} image alignments.}

\begin{figure*}[h]
\includegraphics[scale=0.625,trim = {0 0 0 0}]{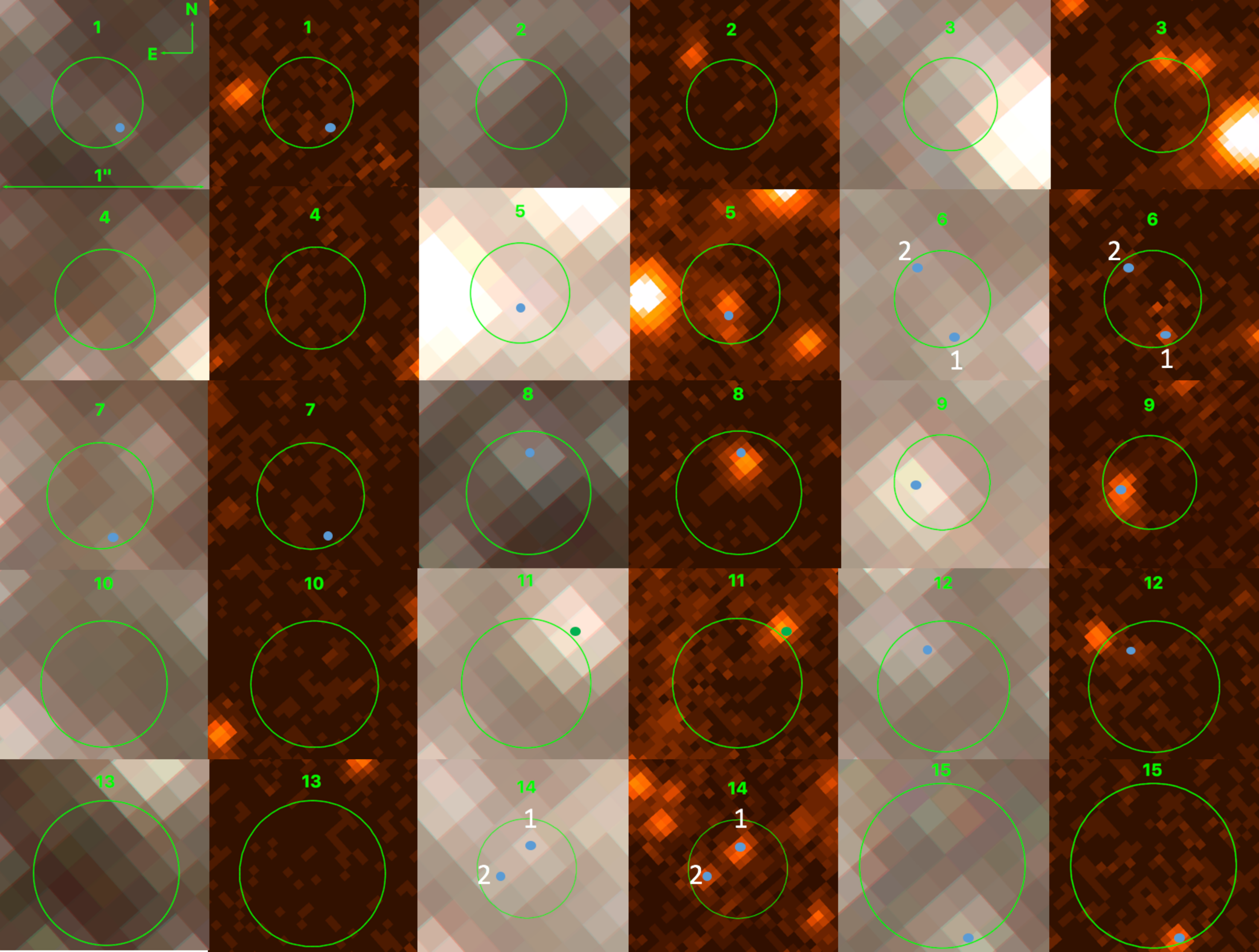}
\caption{WFC3 NIR and UVIS image cutouts of the vicinity of  X-ray sources. North is up and East is to the left. { The numbering is the same as in the left panel of Figure \ref{ROI}.} The 2$\sigma$ X-ray positional error circles are shown in green (see Section \ref{xraysrcs}).  The NIR images (left columns) are false color (F127M -- blue, F139M -- green, and F153M -- red). The UVIS F814W images (right columns) show the same area of the sky. The blue dots represent the positions of the NIR sources detected in the drizzled WFC3/IR F127M image. In cases with multiple potential NIR counterparts they are numbered (the numbers are used as subscripts to the source names in Table \ref{counterparts}). The green dot shows the potential NIR counterpart to the variable source X11, located just outside of the 2$\sigma$ X-ray position error circle. { The {\sl James Webb Space Telescope's} ({\sl JWST}) NIRCam will have an angular resolution comparable to that of WFC3/UVIS at NIR wavelengths (see \url{https://jwst-docs.stsci.edu/display/JTI/NIRCam+Imaging}).}
}
\label{cutout}
\end{figure*}

\section{Data Analysis}
\label{analysis}

\subsection{X-ray Sources}
\label{xraysrcs}
{ The source detection procedure (see Section \ref{cacis} for details) found 15  X-ray sources with $\geq$6 net counts within $42''$ (see Section \ref{hstphot}) of the cluster center reported by \cite{2005AJ....129..239K}. Each sources X-ray properties were extracted using ACIS Extract, which also calculates the {\em p}-value ($P_B$) for the source extraction under the no-source null hypothesis (i.e., that all counts in the source aperture are background; \citealt{2010ApJ...714.1582B})}. Out of the 15 X-ray sources numbered in Figure \ref{ROI} (see left panel), source X8 has the largest $P_B$=7$\times10^{-5}$ in the 0.5--8 keV energy band. Therefore, all of these sources are confidently detected. The  uncertainties of the X-ray source positions are computed from the statistical errors (from centroiding) and systematic errors from the absolute astrometry of {\sl CXO} and {\sl HST} (see Section \ref{astrometry}) added in quadrature. We consider the 2$\sigma$ combined statistical and systematic positional uncertainties of the X-ray sources when searching for optical/NIR counterparts. The mean 2$\sigma$ positional uncertainty is 0\farcs28.

For each source, we calculated an unabsorbed X-ray color defined as $X_{\rm color}=\log({F^{\rm unabs}_{\rm 0.5-2 keV}/F^{\rm unabs}_{\rm 2-6 keV}}$), as well as the unabsorbed X-ray luminosity in the 0.5$-$8 keV band (see Table \ref{xsrc}). The unabsorbed fluxes are calculated using the best-fit photon index ($\Gamma$) and the $N_{\rm H}$ adopted above. All fainter sources, that do not pass the number of counts threshold  for spectral fitting,  have median photon energies $>2.5$ keV, suggesting rather hard spectra (see Table \ref{xsrc}). Therefore, to calculate the unabsorbed flux and luminosity for these sources we assume $\Gamma=1.0$, which is the most typical best-fit value obtained for the brighter sources  (see Table \ref{xsrc}). { Previously, \cite{2007arXiv0708.3365P} used a hydrogen absorption column density $N_{\rm H}=2.7\times10^{22}$ cm$^{-2}$ in the absorbed power-law fits to the spectra. This $N_{\rm H}$ was derived from the $N_{\rm H}-A_V$ relation of \cite{1995A&A...293..889P} using an $A_V=15$ reported by \cite{2005AJ....129..239K}. For comparison, we have refit our spectra for sources X1-X6 with the the same $N_{\rm H}$ as \cite{2007arXiv0708.3365P} and found photon indices within 1.5$\sigma$ of the uncertainties of our fits\footnote{\cite{2007arXiv0708.3365P}  do not provide the uncertainties for their fits.}.}

Extended emission clearly fills most of the region within the $r=36''$ half-light radius (see Figure \ref{ROI}). After removing all confidently detected point sources with $>6$ net counts, the remaining net count rate within this radius is $0.012\pm0.002$ counts s$^{-1}$. However, this emission may not be truly diffuse, but rather composed of faint unresolved point sources. { The unresolved background emission can be fit by an absorbed power-law model, with a hydrogen absorption column density $N_{\rm H}=2.9^{+0.7}_{-0.6}\times10^{22}$ cm$^{-2}$ and photon index $\Gamma=1.8\pm0.4$, corresponding to an observed luminosity of $L_{X}\approx3\times10^{32}$ erg s$^{-1}$ in the 0.5-8 keV band. If, instead, we freeze the hydrogen absorption column density to N$_{\rm H}=4\times10^{22}$ cm$^{-2}$, the photon index becomes $\Gamma=2.4\pm0.2$.}

The faintest source in Table \ref{xsrc} has an absorbed limiting flux of 4$\times10^{-15}$ erg cm$^{-2}$ s$^{-1}$ in the 0.5--8 keV energy band. This corresponds to an unabsorbed luminosity limit of 8$\times10^{30}$ erg s$^{-1}$ at a plausible distance of 3.3 kpc (see Section \ref{agediscuss}), assuming an absorbed power-law model with a photon index $\Gamma=1$ and hydrogen absorption column N$_{\rm H}=4\times10^{22}$ cm$^{-2}$. { Deeper {\sl CXO} observations are {\sl planned} to probe the fainter end of the X-ray source population in GC01.}

\subsection{Optical and NIR Photometry with HST}
\label{hstphot}

To select sources with reliable photometry from the DOLPHOT catalog  several quality cuts must be made. The most important quality control parameters for the photometric catalog are the signal-to-noise ratio (SNR), the crowding parameter, the sharpness parameter, and a photometric flag.  Only sources with ``Photometric Flag''$=0$ (i.e., this ensures that sources near the image edge or those with saturated pixels are removed) in each filter were kept for the photometry. The ``crowd'' parameter shows how much brighter (in magnitude) a star would be if other nearby stars were not fit simultaneously, while the ``sharp'' parameter describes how peaked, compared to the PSF, a source is$^{\ref{note1}}$.

GC01 has both substantial crowding, which affects the photometry and a large and variable { reddening} (see Section \ref{colcol}). WFC3/UVIS has a higher spatial resolution than WFC3/IR (0\farcs04 versus 0\farcs13 pixel scales, respectively), making it better at resolving blended sources. However, UVIS is not able to detect as many sources as WFC3/IR due to the large { reddening}. For this reason we ran DOLPHOT twice, first using the drizzled F127M image and then using the drizzled F814W image for source detection. Once source detection is complete, DOLPHOT extracts the photometry from the raw frames in each WFC3 filter. Our analysis below focuses primarily on the photometry extracted using the WFC3/IR F127M image for source finding. This is due to the fact that a number of WFC3/IR counterparts to X-ray sources do not have a WFC3/UVIS counterpart. Below, we will explicitly state when the WFC3/UVIS F814W image has been used for source finding.

For the NIR photometry, in addition to requiring ``Photometric Flag''$=0$, we removed all sources with SNR$<4$, {\tt crowd}$>0.2$ and {\tt sharp}$>$0.03 in all WFC3/IR images. { We chose all sources within r$=42''$ of the cluster center determined by \cite{2005AJ....129..239K}}. The size of the region was chosen to analyze more sources belonging to the cluster, and, at the same time, to avoid the dark patchy areas appearing to the west of the cluster. Below, this region is  referred to as the region of interest (ROI). After the cuts to the NIR photometry, 1,964 sources remained within the ROI. 

For the UVIS/NIR photometry using the WFC3/UVIS F814W image for source detection, we used the same ROI and IR cuts listed above. However, to retain more sources the cuts were only applied to the F127M image. We also cut all sources with a UVIS SNR $<4$, {\tt crowd} $>0.2$, and {\tt sharp} $>$0.1 leaving { 777} UVIS/NIR sources.

\subsection{Optical and NIR Counterparts to X-ray Sources}
\label{chance}
The confidences of the NIR source detections may be significantly impacted by the unaccounted for systematic uncertainties in the background measurements due to the large amount of crowding in this field. Therefore, to ensure that the NIR counterparts to X-ray sources were confidently detected, we required that they have SNR$>30$ in each of the three WFC3/IR filters. Any source meeting the SNR$>30$ requirement in all of the F127M, F139M, and F153M images, and located within the 2$\sigma$  error circle of an  X-ray source listed in Table \ref{xsrc} is considered to be a potential counterpart to the corresponding X-ray source (see Figure \ref{cutout} and Table \ref{counterparts}). The only exception is made for source X11, because it is highly variable in X-rays and has a SNR$>$30 counterpart just outside of its error circle. Most of the considered X-ray sources are located in the  crowded core of GC01, meaning that their NIR counterparts have large crowding values (i.e., $>0.2$), with the exceptions of X8, X9, and X11. In Table \ref{counterparts}, we only provide F814W photometry for the counterparts of NIR sources detected with SNR$>$10 in the F814W image. The lower threshold for the SNR is enabled by the higher resolution { and smaller number of sources (i.e., leading to less crowding) in} the F814W image.

We find that  { nine} X-ray sources have at least one NIR counterpart, while three of these X-ray sources (X5, X6, X14) have two counterpart candidates within their X-ray error circles (see Figure \ref{cutout}). For these three sources, the { potential} NIR counterparts are referred to by using a subscript after the X-ray source name (e.g., X6$_1$, X6$_2$). Since we used the NIR image as the source detection image, the two NIR sources within the X-ray error circle of source X5 were not resolved and were considered a single NIR source (although they are resolved in the F814W image).  

\begin{deluxetable*}{lcccccccccccc}
\tablecaption{X-ray sources detected with { $\geq$} 6 net counts in the ACIS-S3 image within the $42''$ ROI.\label{xsrc}}
\tablehead{
\colhead{Src} & \colhead{R.A.} & \colhead{Decl.} & $\delta r$\tablenotemark{a} & \colhead{Net Cts} & \colhead{$F_X$\tablenotemark{b}} &  \colhead{$L_X$\tablenotemark{c}} & \colhead{$X_{\rm col}$\tablenotemark{d}} & \colhead{$E_{\rm med}$\tablenotemark{e}}  & \colhead{$\Gamma$\tablenotemark{f}} & \colhead{C-stat/d.o.f.}  &  ${ P_{\rm NIR}\tablenotemark{j}}$ & ${ P_{\rm opt}\tablenotemark{k}}$
 }
 \startdata
 X1 & 282.209462 & $-$1.491307 & 0.23 & 61$\pm8$ & 3.6$\pm0.5$ & 9.1 & -0.1 & 4.1 & 0.9$\pm0.4$ & 67.00/58 & { 39\%} & { 11\%}\\
 X2 & 282.203688 & $-$1.499904 & 0.22 & 52$\pm7$ & 1.2$^{+0.3}_{-0.2}$ & 340 & 2.1 & 1.8 & 5.6$\pm0.6$ & 24.44/42 & { 40\%} & { 12\%}\\
 X3 & 282.205215 & $-$1.500029 & 0.23 & 39$^{+7}_{-6}$ & 2.4$\pm0.4$ & 4.9  & -0.3 & 4.2 & 0.7$\pm0.4$ & 53.54/39 & { 47\%} & { 18\%}\\
 X4 & 282.207574 & $-$1.491672 & 0.25 & 34$\pm6$ & 2.2$\pm0.4$ & 3.7 & -0.8 & 4.4 & $-$0.2$\pm0.6$ & 34.70/32 & { 45\%} & { 13\%}\\
 X5 & 282.208201 & $-$1.496848 & 0.25  & 31$^{+6}_{-5}$ & 2.0$\pm0.4$ & 4.4 & -0.3 & 4.2 & 1.2$\pm0.5$ & 29.97/28 & { 65\%} & { 37\%}\\
 X6 & 282.207103 & $-$1.499042 & 0.24  & 30$^{+6}_{-5}$ & 1.8$^{+0.4}_{-0.3}$ & 4.5 & -0.1 & 4.0 & 1.1$\pm0.6$ & 37.07/25 & { 59\%} & { 31\%}\\
 X7 & 282.206333 & $-$1.494469 & 0.27  & 20$^{+5}_{-4}$ & 1.3$\pm0.3$ & 2.8  & ...\tablenotemark{g}  & 4.3 & 1.0\tablenotemark{h} & ... & { 59\%} & { 25\%}\\
 X8 & 282.216777 & $-$1.494315 & 0.31 & 14$\pm4$ & 0.6$\pm0.2$ & 1.7 & -0.1 & 3.4 & 1.0\tablenotemark{h} & ... & { 47\%} & { 11\%}\\
 X9 & 282.209069 & $-$1.498165 & 0.23  & 10$^{+4}_{-3}$ & 0.8$\pm0.3$ & 3.0  & 0.4 & 2.6 & 1.0\tablenotemark{h} & ... & { 59\%} & { 32\%}\\
X10 & 282.208920 & $-$1.500203  & 0.31 & 9$^{+4}_{-3}$ & 0.4$^{+0.2}_{-0.1}$ & 0.9 & ...\tablenotemark{g} & 3.2 & 1.0\tablenotemark{h} & ... & {76\%} & { 44\%}\\
X11 & 282.206564 & $-$1.496531 & 0.32  & 9$^{+4}_{-3}$ & 0.5$\pm0.2$\tablenotemark{i} & 1.1\tablenotemark{i} & -0.3 & 3.8 & 1.0\tablenotemark{h} & ... & {83\%\tablenotemark{ {l}}} & { 51\%\tablenotemark{ {l}}}\\
X12 & 282.206393 & $-$1.499782 & 0.32  & 8$\pm3$ & 0.4$^{+0.2}_{-0.1}$ & 0.9 & ...\tablenotemark{g} & 3.4 & 1.0\tablenotemark{h} & ... & {74\%} & { 38\%}\\
X13 & 282.199781 & $-$1.495440 & 0.36  & 7$\pm3$ & 0.4$^{+0.2}_{-0.1}$ & 0.8 & ...\tablenotemark{g} & 3.8 & 1.0\tablenotemark{h} & ... & {54\%} & { 12\%}\\
X14 & 282.208978 & $-$1.498018 & 0.24 & 7$\pm3$ & 0.8$^{+0.4}_{-0.3}$ & 1.7 & ...\tablenotemark{g} & 3.1 & 1.0\tablenotemark{h} & ... & {62\%} & { 35\%}\\
X15 & 282.213183 & $-$1.495756 & 0.40  & 6$^{+3}_{-2}$ & 0.4$^{+0.2}_{-0.1}$  & 0.8 & ...\tablenotemark{g} & 3.8 & 1.0\tablenotemark{h} & ...  & {84\%} & { 42\%}\\
\enddata
\tablenotetext{\rm a}{X-ray 2$\sigma$ position error circle radius in arcseconds.}
\tablenotetext{\rm b}{Observed 0.5--8\,keV flux in units of $10^{-14}$\,erg\,cm$^{-2}$ s$^{-1}$.}
\tablenotetext{\rm c}{Unabsorbed 0.5--8\,keV luminosity in units of $10^{31}$\,erg\,s$^{-1}$ at $d=$3.3 kpc.}
\tablenotetext{\rm d}{X-ray color defined as $X_{\rm color}=\log{(F^{\rm unabs}_{\rm 0.5-2 keV}/F^{\rm unabs}_{\rm 2-6 keV})}$.}
\tablenotetext{\rm e}{Median photon energy in the 0.5--8 keV band.}
\tablenotetext{\rm f}{Best fit photon index for the absorbed power-law model.}
\tablenotetext{\rm g}{Sources have no counts detected in the soft (0.5-2.0 keV) band. }
\tablenotetext{\rm h}{Photon index set to $\Gamma=$1.0 for sources with $<$ 30 counts. }
\tablenotetext{\rm i}{Average observed flux and unabsorbed luminosity. { During the 200 s flare the source reached a peak flux $F_X=1\times10^{-12}$ erg cm$^{-2}$ s$^{-1}$ and unabsorbed luminosity $L_X=2.1\times10^{33}$ erg s$^{-1}$} in the 0.5--8\,keV band, assuming $d=$3.3 kpc (see Section \ref{agediscuss})} 
\tablenotetext{\rm j}{Probability that the X-ray source has 1 or more NIR counterparts within its  2$\sigma$ positional error circle due to chance (see Section \ref{chance}).}
\tablenotetext{\rm k}{{ Probability that the X-ray source has 1 or more optical+NIR counterparts within its  2$\sigma$ positional error circle due to chance (see Section \ref{chance}).}}
\tablenotetext{\rm l}{{ The NIR sources is just outside of the $2\sigma$ positional error circle of this source, so we adopt the distance to the NIR source $(\delta r=0\farcs35$) when calculating the chance coincidence probability.}}
\end{deluxetable*}

{ The fact that several X-ray sources have two NIR counterparts is not surprising given the high density of NIR sources in the cluster's core. We estimate the chance coincidence probabilities of finding one or more NIR sources within the 2$\sigma$ positional error circles of each of the X-ray sources. 
 To accomplish this, we first apply the same cuts to the photometric catalog as for the counterpart matching (i.e., SNR$>30$ in all WFC3/IR filters). Next, we divide the $r<42''$ ROI into seven 6$''$-wide concentric annuli because the source density changes as a function of distance from the cluster center. Then, in each annulus, we calculate the source density, $\rho$, of NIR sources.  Finally, we calculate the probability of having one or more NIR sources within randomly placed circles of radii equal to the 2$\sigma$ source positional uncertainties, $P_{\rm NIR}=1-\exp(-\rho_{\rm NIR} \pi \delta r_{2\sigma}^{2})$. We then repeated this exercise using sources that pass both the NIR and UVIS cuts (SNR$>30$ and SNR$>10$, respectively). The chance coincidence probabilities ($P_{\rm NIR}$ and $P_{\rm opt}$) for each X-ray source are shown in the last column of Table \ref{xsrc}.

Out of the 15 X-ray sources listed in Table \ref{xsrc}, 9$\pm2$ are expected to be coincident with NIR sources by chance, while 4$\pm2$ are expected to be coincident with a NIR$+$optical source by chance. Although the expected number of chance coincident NIR sources is close to the number of X-ray sources with NIR counterparts (9 or 10 if X11 is counted), the number of coincidences between the NIR$+$optical sources is about two times larger than is expected by random chance. This suggest that at least some of the counterparts are likely to be real\footnote{{ Extreme crowding could  lead to unaccounted for systematic errors in the Dolphot PSF fitting procedure and some of the faint NIR sources, particularly those in the vicinity of brighter NIR sources, could be spurious leading to an overestimated density of NIR sources and, consequently, an overestimated $P_{\rm NIR}$}.}. 

Deeper X-ray and NIR observations, or spectroscopic observations of the NIR sources coincident with the X-ray sources, can help to establish a physical relationship between the X-ray and NIR/optical sources.}



\begin{figure*}
\includegraphics[scale=0.6]{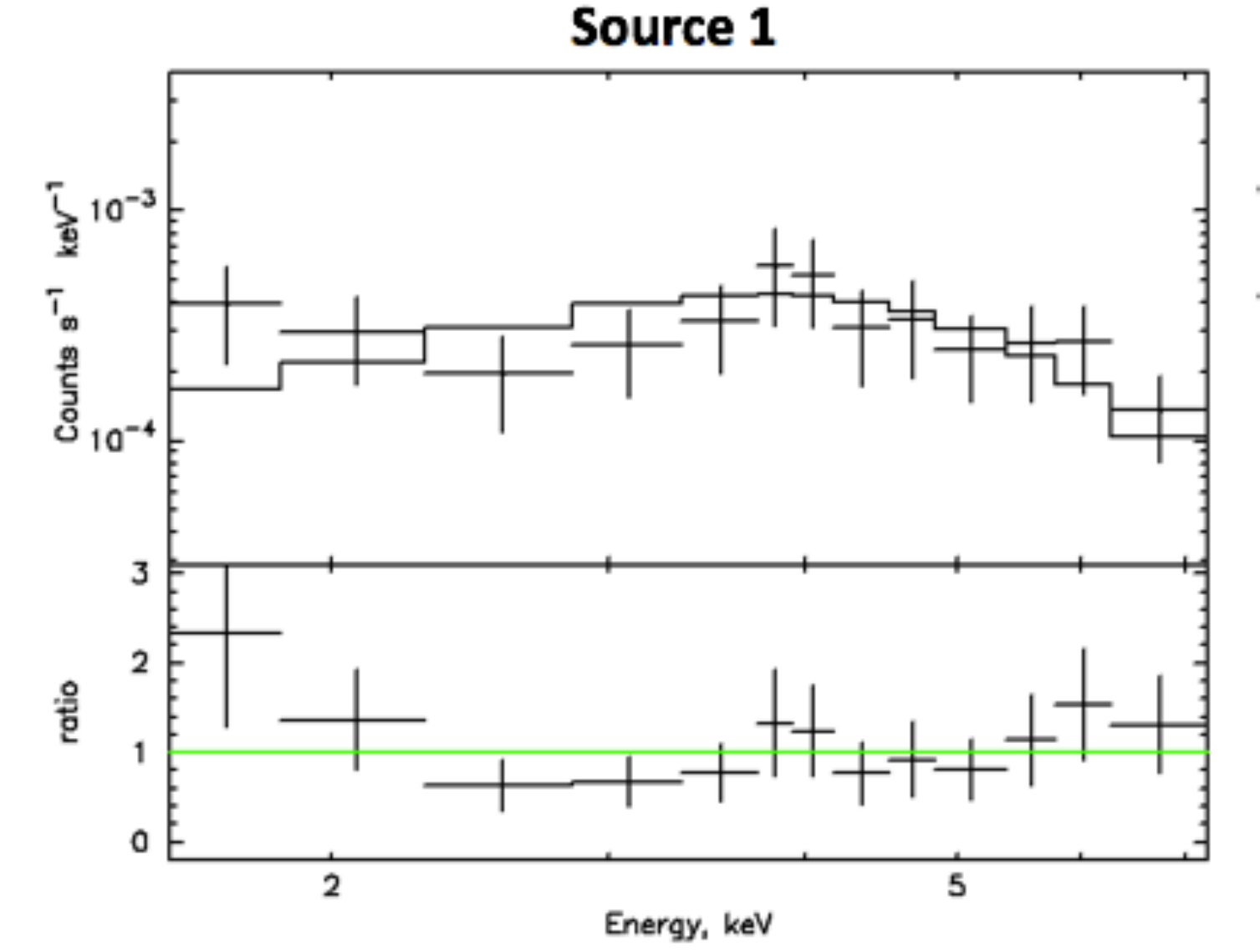}
\includegraphics[scale=0.6]{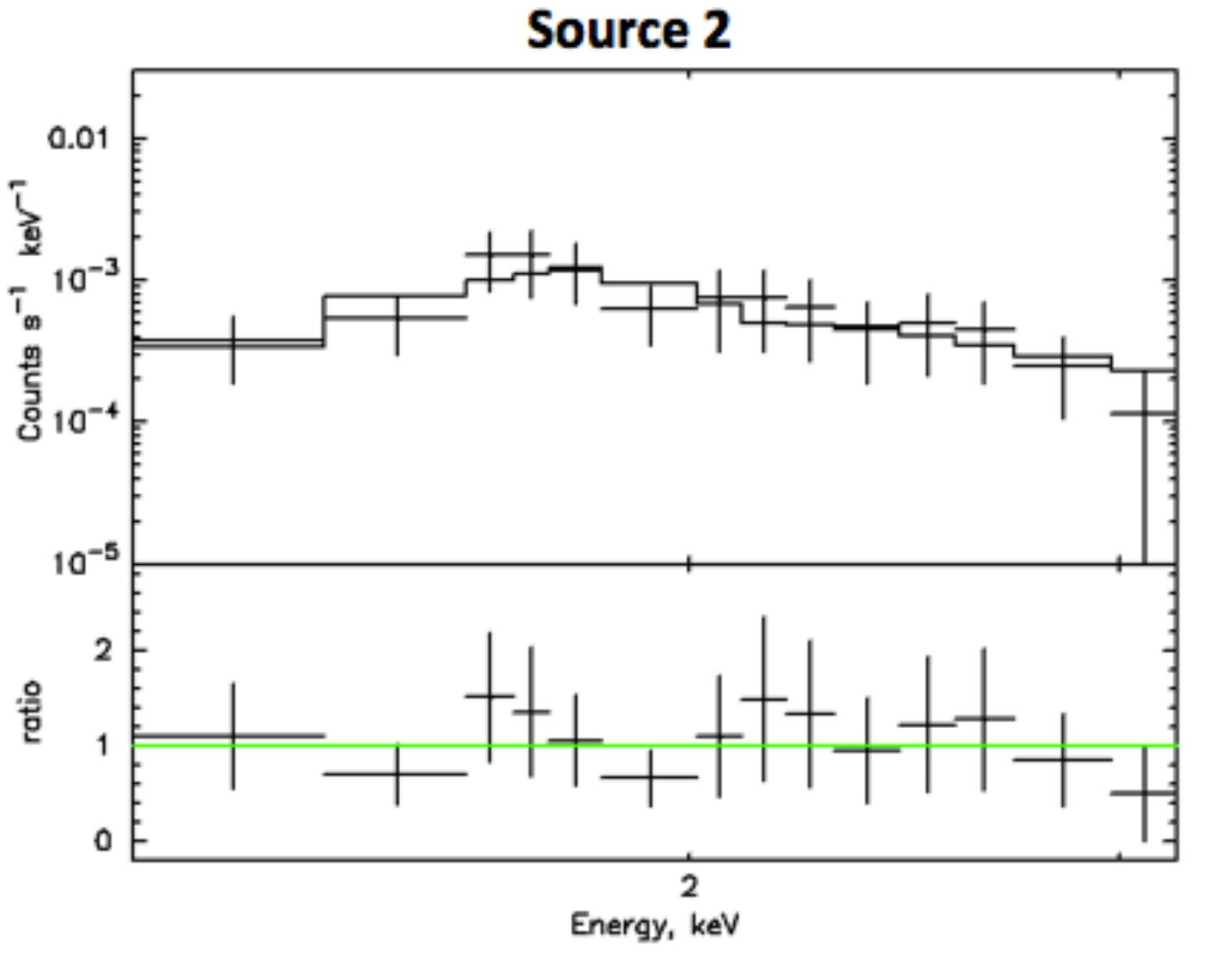}
\includegraphics[scale=0.6]{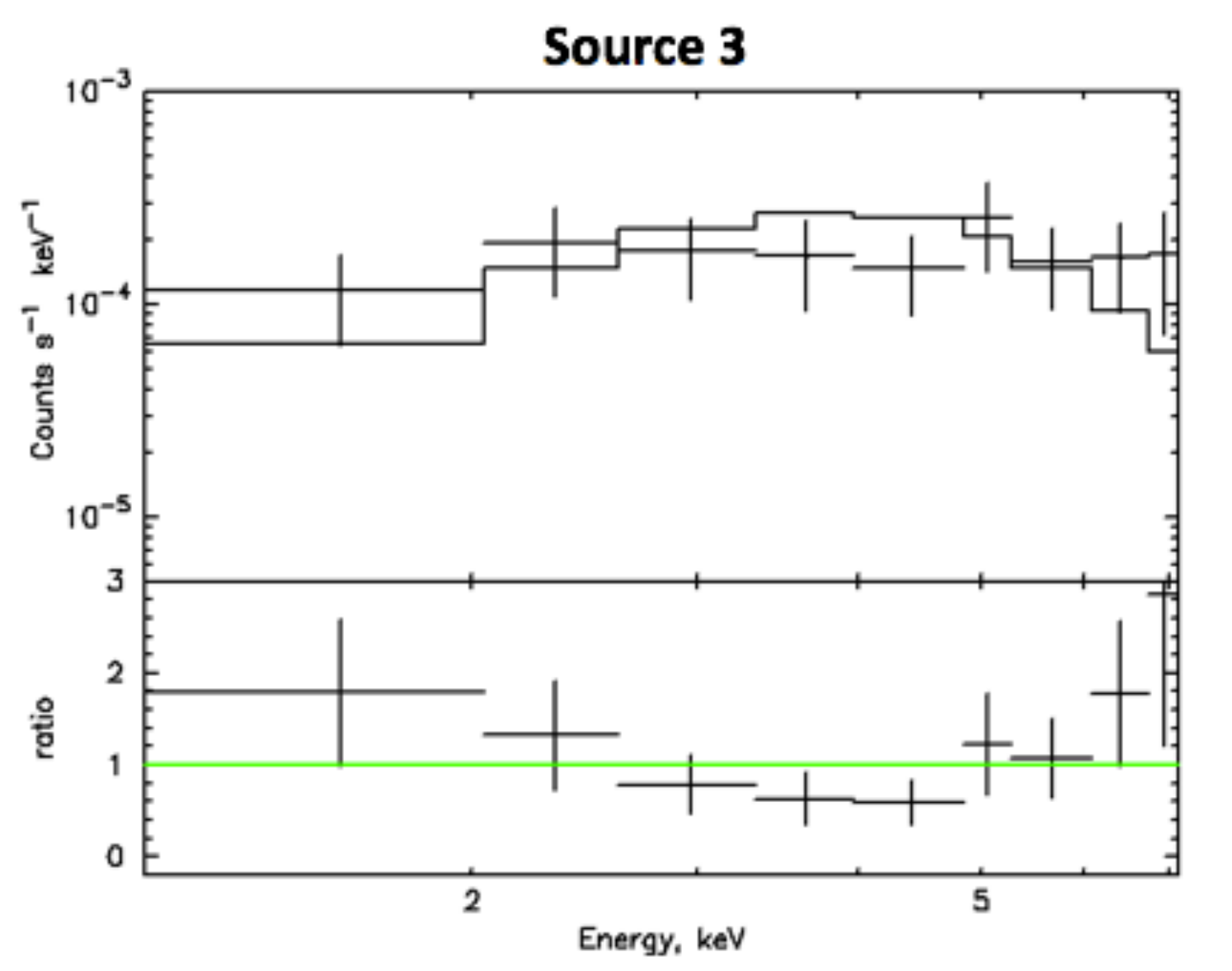}
\includegraphics[scale=0.6]{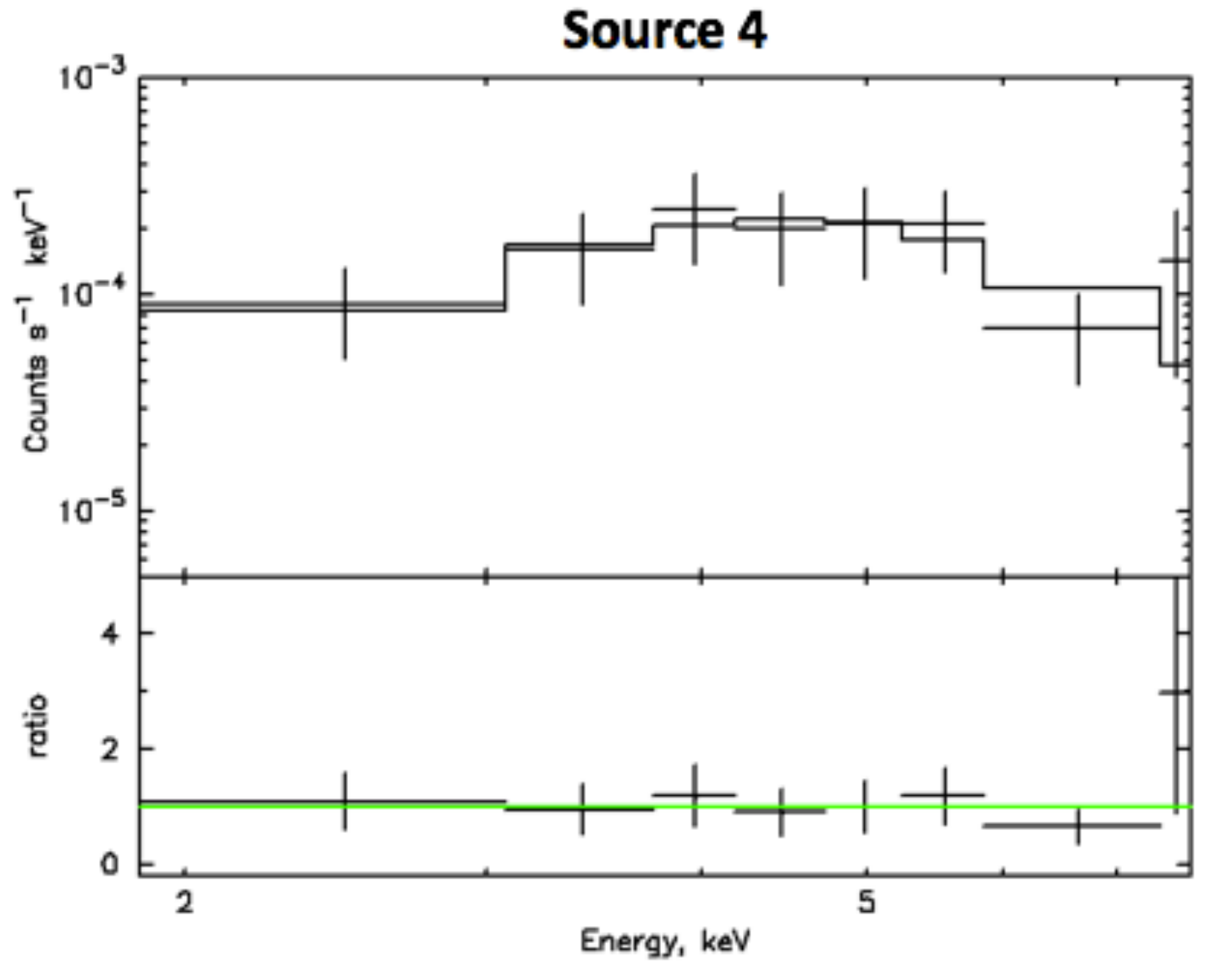}
\includegraphics[scale=0.6]{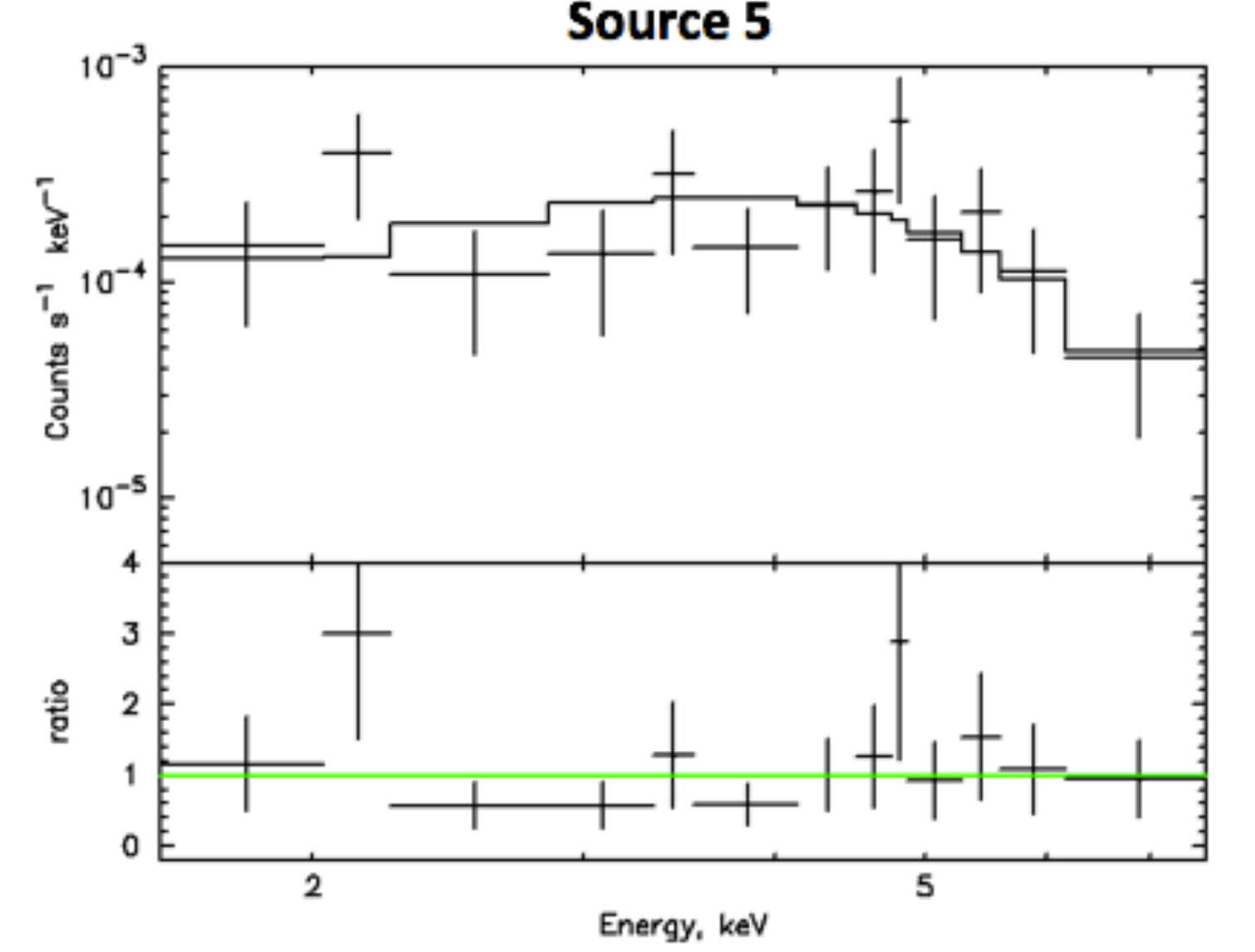}
\includegraphics[scale=0.6]{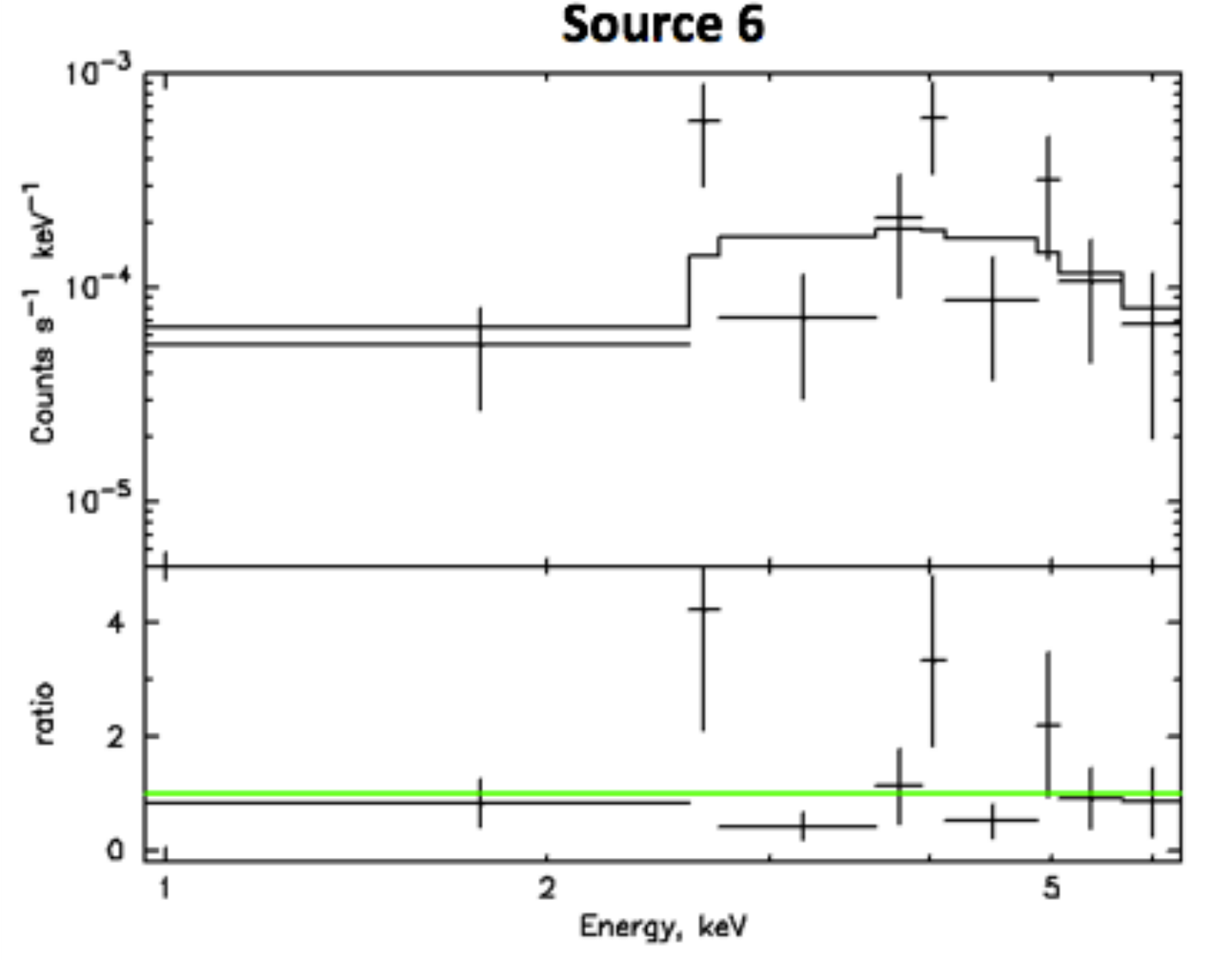}
\caption{X-ray  spectra for six sources detected with $>30$ counts. All sources are fit with an absorbed power-law model with fixed N$_{\rm H}=4.0\times10^{22}$ cm$^{-2}$ (see Section \ref{cacis}). The fits were performed using Cash statistics (c-stat) using unbinned data (the spectra are binned for visualization purposes only). The best-fit parameters are given in Table \ref{xsrc}.  }
\label{xspec}
\end{figure*}

\section{Properties of the GLIMPSE-C01 Cluster}

In this section we determine the position of the cluster's center and then use the photometry to produce the color-color diagram, CMDs, and set constraints on GC01's distance, age, and extinction. All isochrones are adopted from \cite{2017ApJ...835...77M} and are re-calculated for the three {\sl HST} WFC3/IR filters and the WFC3/UVIS filter\footnote{See \url{http://stev.oapd.inaf.it/cgi-bin/cmd} for details.}. Finally, we show that the puzzling `looplike structure'' reported by \cite{2005AJ....129..239K} is comprised of stars unresolved in shallower survey images.

\begin{deluxetable*}{lcccccccccc}
\tablecaption{Potential NIR counterparts to X-ray sources and their magnitudes in the WFC3 IR and UVIS filters (see Section \ref{chance} for details).\label{counterparts}}
\tablehead{
\colhead{Source} & \colhead{RA} & \colhead{DEC} & \colhead{$m_{F814W}$\tablenotemark{a}} &  \colhead{$\Delta m_{F814W}$\tablenotemark{b}} & \colhead{$m_{F127M}$} &  \colhead{$\Delta m_{127M}$\tablenotemark{b}} & \colhead{$m_{F139M}$} &  \colhead{$\Delta m_{F139M}$\tablenotemark{b}} & \colhead{$m_{F153M}$} &  \colhead{$\Delta m_{F153M}$\tablenotemark{b}} 
}
\startdata
 X1 & 282.209434 & -1.491339 &  ... &  ... &  21.184 &  0.032 &  20.383 &  0.022 &  19.556 &  0.016 \\
 X5\tablenotemark{c} & 282.208204 & -1.496877 &  24.229 &  0.070 &  17.228 &  0.004 &  16.530 &  0.003 &  15.852 &  0.003 \\
   X6$_1$ & 282.207082 & -1.499093 &  24.598 &  0.092 &  18.284 &  0.006 &  17.554 &  0.005 &  16.826 &  0.004 \\
   X6$_2$ & 282.207137 & -1.498999 &  ... &  ... &  19.135 &  0.010 &  18.439 &  0.008 &  17.764 &  0.006 \\
   X7 & 282.206317 & -1.494525 &  ... &  ... &  20.597 &  0.024 &  19.587 &  0.015 &  18.699 &  0.011 \\
   X8 & 282.216775 & -1.494257 &  23.998 &  0.059 &  19.242 &  0.010 &  18.649 &  0.008 &  18.035 &  0.006 \\
   X9\tablenotemark{{$\dagger$}} & 282.209104 & -1.498178 &  23.435 &  0.041 &  16.478 &  0.002 &  15.652 &  0.002 &  14.848 &  0.001 \\
 X11\tablenotemark{{d,$\dagger$}} & 282.206500 & -1.496457 &  23.314 &  0.039 &  16.486 &  0.002 &  15.646 &  0.002 &  14.832 &  0.001 \\
 X12 & 282.206414 & -1.499733 & ... &  ... &  19.477 &  0.013 &  18.792 &  0.011 &  18.141 &  0.009 \\
  X14$_1$ & 282.208970 & -1.497989 &  24.099 &  0.063 &  18.102 &  0.006 &  17.419 &  0.005 &  16.800 &  0.004 \\
  X14$_2$ & 282.209016 & -1.498029 &  24.527 &  0.085 &  18.909 &  0.009 &  18.272 &  0.008 &  17.676 &  0.007 \\
    X15 & 282.213147 & -1.495857 &  23.809 &  0.052 &  18.078 &  0.005 &  17.387 &  0.004 &  16.687 &  0.003 \\
\enddata
\tablenotetext{\rm a}{UVIS magnitude for sources detected with SNR$>10$.}
\tablenotetext{\rm b}{The magnitude uncertainties are statistical only and do not include systematic errors due to crowding.}
\tablenotetext{\rm c}{The source is resolved into two individual sources in the higher resolution UVIS image.}
\tablenotetext{\rm d}{This NIR/optical source lies just outside of the X-ray positional error circle. However, due to the highly variable nature of this X-ray source, we consider it as a potential NIR/Optical counterpart.}
\tablenotemark{$\dagger$}{Unlike other sources, this source meets the criteria to be included in the photometric catalog (i.e., SNR$>$4, {\tt crowd}$<$0.2, and {\tt sharp}$<$0.03).}
\tablecomments{All magnitude are in the VEGAMAG system.}
\end{deluxetable*}

\label{clusprop}
\subsection{Cluster Center and Size}
\label{cent}

{ We attempted to determine the cluster's center by fitting ellipses to the isophotal contours (isophotes) of the cluster using the drizzled WRC3/IR F127M filter image. However, the cluster center determination is affected by the dust lanes to the west of the cluster (see Figure \ref{ROI}). Figure \ref{spitzer} shows the contours from the CO High-Resolution Survey (COHRS; \citealt{2013ApJS..209....8D}). We find strong CO emission on distances as close as $\sim10''$ to the cluster center, prohibiting an accurate determination of the cluster's center from the {\sl HST} data. For these reasons we adopt the cluster center provided in \cite{2005AJ....129..239K} (i.e., RA(J2000)$=282.20708^{\circ}$, DEC(J2000)$=-1.49722^{\circ}$), who relied on {\sl Spitzer} IRAC images, which are much less affected by reddening.}

We did not attempt to calculate the half-light radius of GC01 because the relatively small field of view of the WFC3 detectors prevents an accurate estimation of the background light.  The background measurements can have a large impact on the half-light radius estimate. For instance, \cite{2005AJ....129..239K} reported a half-light radius of 36$''$, while \cite{2011MNRAS.411.1386D} found a half-light radius of 14$''$ and suggested that this difference could come from the estimate of the large amount of background light. In this paper we adopt the half-light radius value of 36$''$ reported by \cite{2005AJ....129..239K}.

\subsection{$A_V$ from Color-Color Diagram}
\label{colcol}
In order to determine the extinction\footnote{We use the extinction parameterization provided in \cite{1999PASP..111...63F}.} in the direction of GC01 we have plotted the F127M--F153M vs. F139M--F153M color-color diagram (see Figure \ref{colorcolor}).   The red star and the  error bar represent the mean and standard deviation of the color-color distribution. The 1$\sigma$ standard deviations are  0.30 for the F127M--F153M and 0.15 for the F139M--F153M colors. To estimate the extinction, we used 500 Myr, 1, 2, and 3 Gyr solar metallicity isochrones, as well as a 10 Gyr low-metallicity (1/20 solar) isochrone with A$_V=0$. The orange cross marks the main-sequence turn-off (MSTO), while the orange point shows where a 2.2 M$_{\odot}$ sub-giant branch star lies in color-color space for a solar metallicity 1 Gyr isochrone. The blue cross marks the MSTO and the blue point shows where a sub-giant branch 0.9 M$_{\odot}$ star lies in color-color space for a 10 Gyr low-metallicity (1/20 solar) isochrone. We find that an  $A_V=18$ (shown by the green arrow in Figure \ref{colorcolor}) provides an adequate representation of the offset between the mean position of the  colors of the cluster stars and  the sub-giant branch of the isochrones. An $A_V=18$ corresponds to an { A$_J=4.5$ according the relation given by \cite{1999PASP..111...63F}}. The total range of extinctions across the ROI covering the standard deviation in the color-color distribution corresponds to A$_V=14$--22.

\begin{figure}
\includegraphics[scale=0.28]{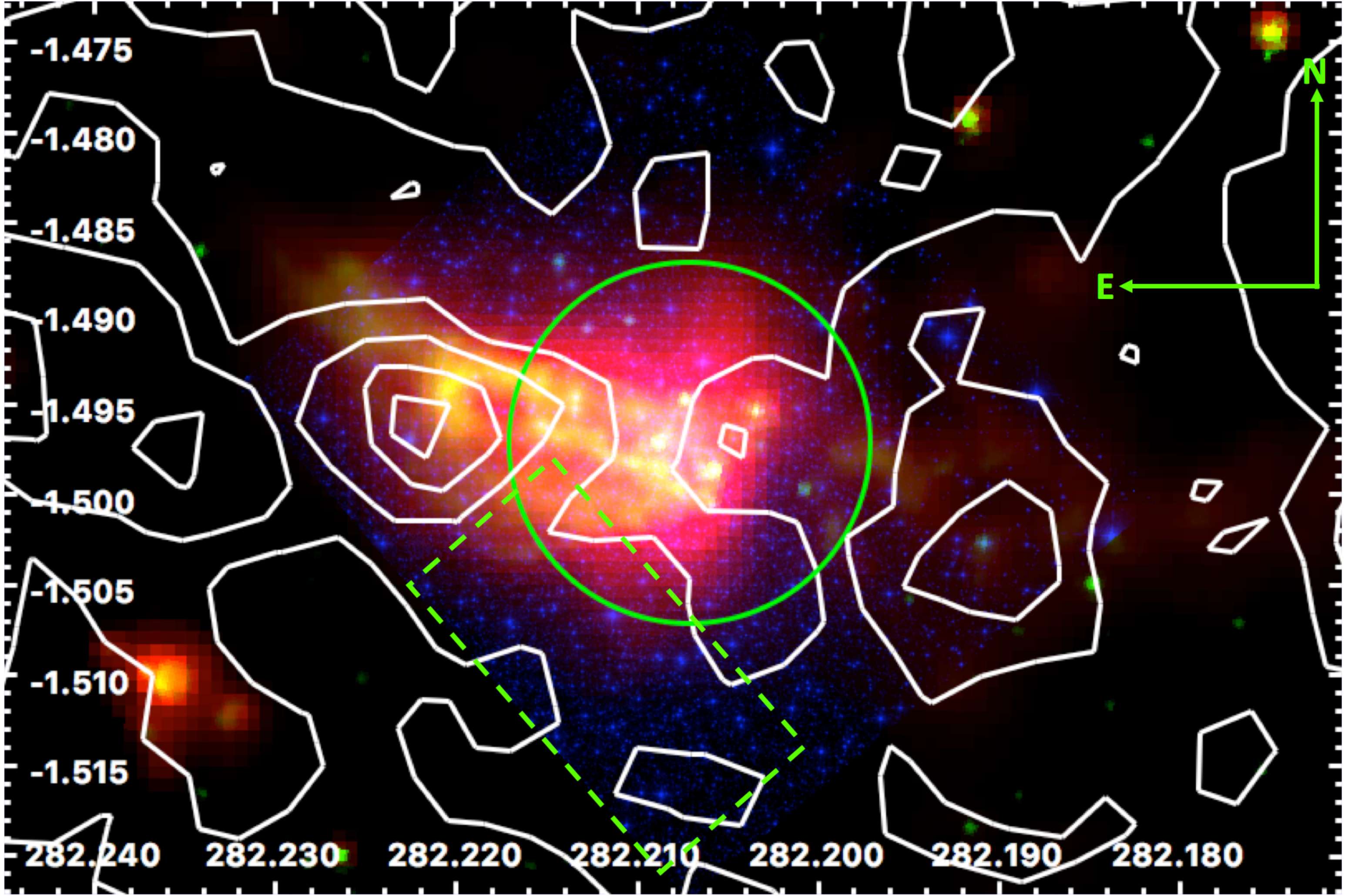}
\caption{Three color image of the GC01 vicinity  from the $24\mu$ \emph{Spitzer} MIPS (red), $8\mu$m \emph{Spitzer} IRAC (green), and F127M filter HST (blue) data (North is up, East is to the left). The $36''$ green circle (half-light radius) is the same as in Figure \ref{ROI}. This image highlights the gas and dust across the cluster, which is the reason for the differential extinction affecting the CMDs (see Section \ref{cmdsec}). { The white contours outline the $^{12}$CO (J$=3\shortrightarrow2$) emission from the CO High-Resolution Survey (COHRS; \citealt{2013ApJS..209....8D}), which heavily overlaps the dust lanes seen to the west of the cluster in Figure \ref{ROI}. The dashed green box shows our alternative CMD extraction region discussed in Section \ref{cmdsec}, which overlaps a relatively uniform region of $^{12}$CO (J$=3\shortrightarrow2$) emission. The equatorial coordinate  grid is shown.}}
\label{spitzer}
\end{figure}

\begin{figure}[h!]
\vspace{0.2cm}
\begin{center}
\includegraphics[scale=0.38, trim={1.5cm 0 0 0}]{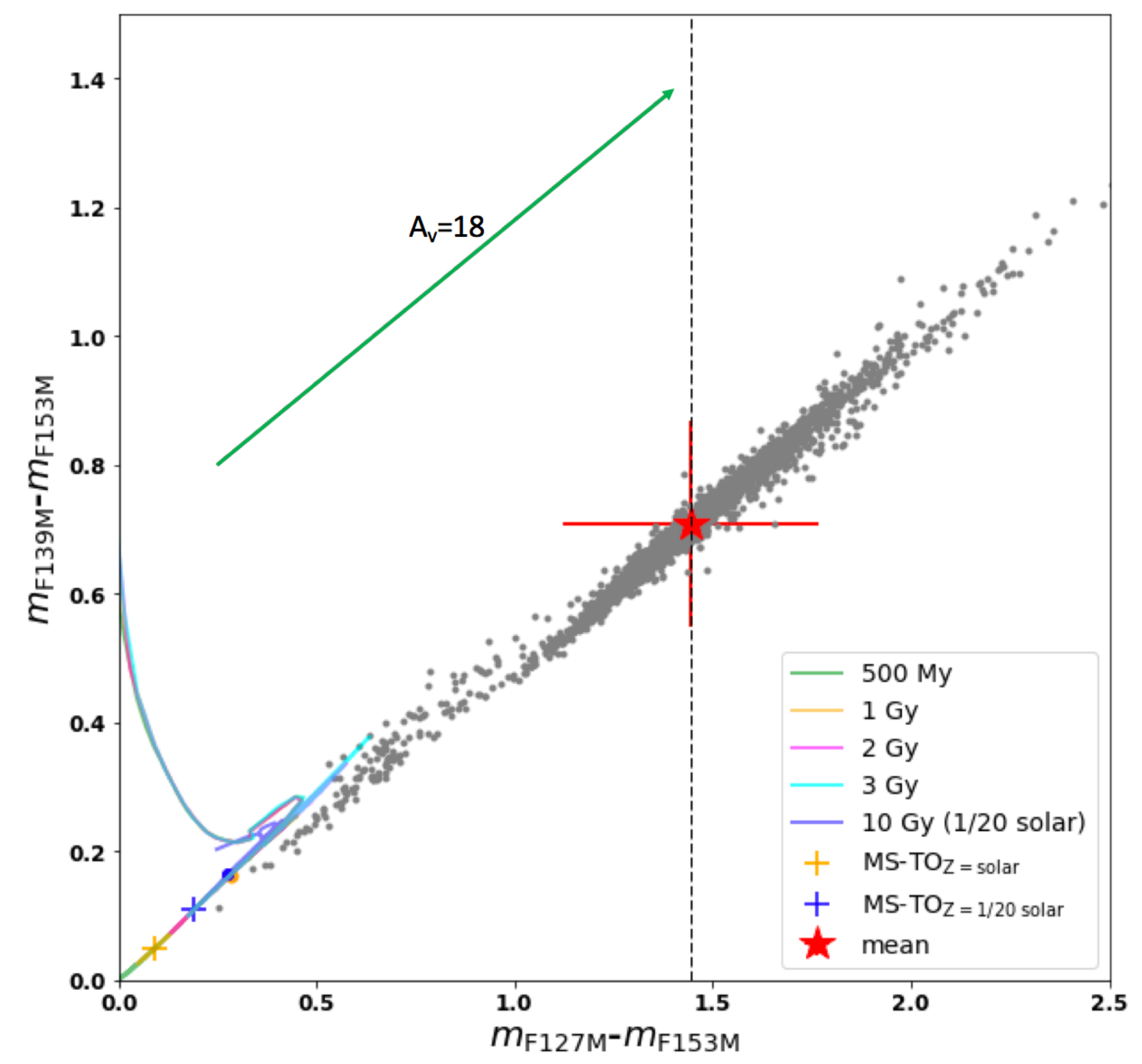}
\caption{IR color-color diagram for all stars within the ROI that passed the quality cuts described in Section \ref{obs}. The 500 Myr, 1 Gyr, 2 Gyr, and 3 Gyr solar metallicity, and the 10 Gyr low-metallicity (1/20 solar) isochrones with A$_V=0$ are shown in green, orange, magenta, cyan, and blue, respectively. The red star shows the mean of the distribution, while the error bars show the 1$\sigma$ standard deviation. The green arrow is the A$_{V}=18$ reddening vector. The orange and blue crosses show the location of the main-sequence turn-off (MSTO) for the 1 and 10 Gyr isochrones, respectively. A 2.2 M$_{\odot}$ star on the 1 Gyr isochrone is indicated by an orange point and a 0.9 M$_{\odot}$ star on the 10 Gyr isochrone is shown as a blue point.}
\label{colorcolor}
\end{center}
\end{figure}

 Our analysis indicates a somewhat higher extinction compared to  A$_V=15\pm3$ found by previous studies (\citeauthor{2005AJ....129..239K} \citeyear{2005AJ....129..239K}; \citeauthor{2005A&A...442..195I} \citeyear{2005A&A...442..195I}). { One possible cause could be the different extinction models used in each study. We use \cite{1999PASP..111...63F}, while \cite{2005AJ....129..239K} and \cite{2005A&A...442..195I} use \cite{1989ApJ...345..245C} and \cite{1985ApJ...288..618R}, respectively. The latter two models predict larger A$_J$ and A$_H$ values  (by up to 10\%) for a given A$_V$ than \cite{1999PASP..111...63F}. In addition, the shallower surveys miss the sources with larger extinctions, leading to a bias in the extinction estimate. A CO molecular cloud in front of the cluster { (see Figure \ref{spitzer} and \citealt{2005AJ....129..239K})}, is  likely to be responsible for most of the extinction, including the differential extinction across the cluster}. { Indeed, the COHRS CO contours and  $24\mu$ \emph{Spitzer} MIPS image (see Figure \ref{spitzer}) clearly show the non-uniform ISM structure in the direction toward GC01. }

\subsection{Red Clump Stars}
\label{redcl}

\begin{figure}
\includegraphics[scale=0.35]{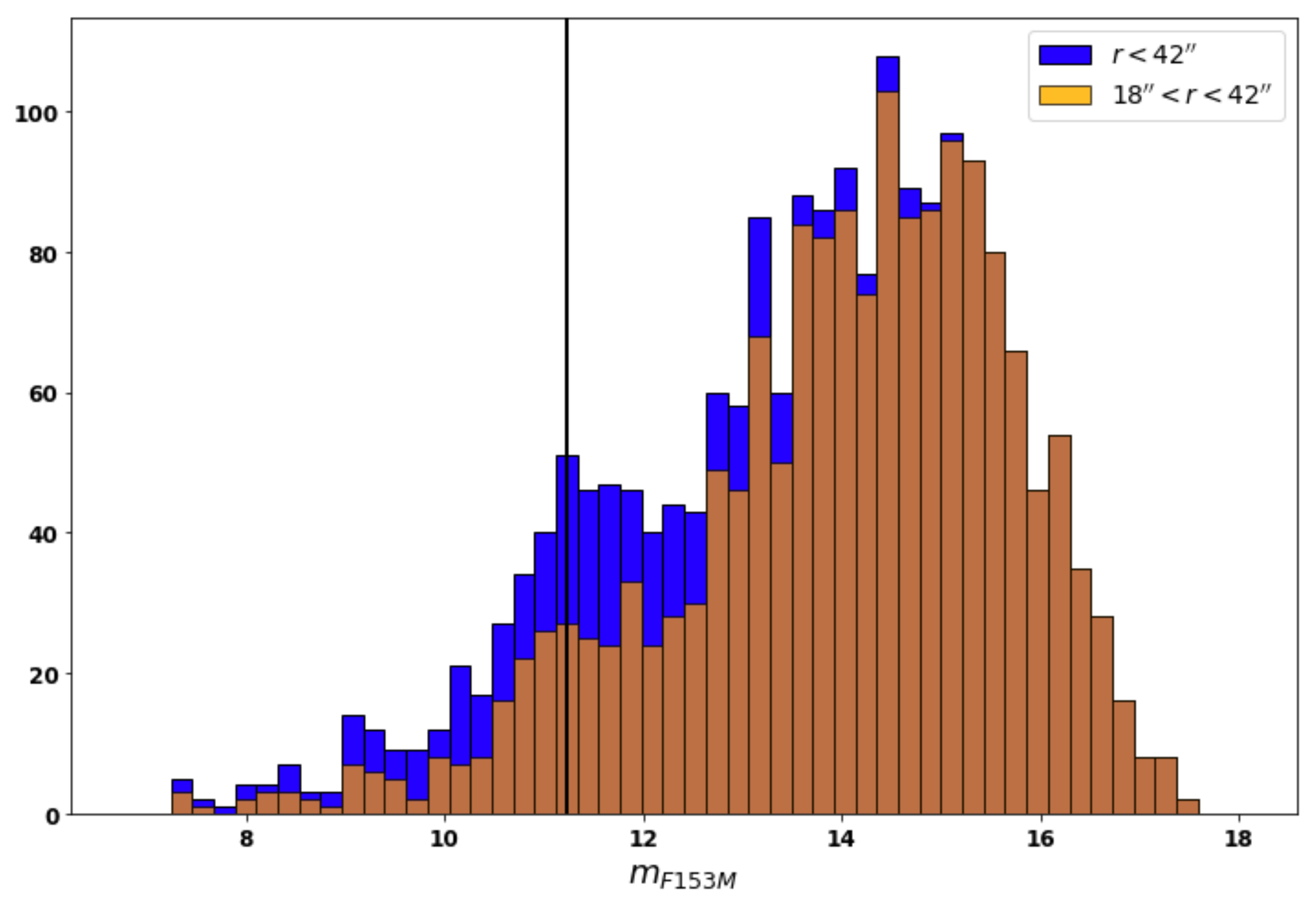}
\caption{{\sl HST} WFC3 F153M luminosity functions for all stars within the cluster ROI (blue) and excluding the cluster core $18''<r<42''$ (orange) with a 0.2 magnitude binning. The luminosity functions are extinction corrected for A$_V=18.0$ (see Section \ref{colcol}). The peak of the red clump stellar distribution at $m_{F153M}={ 11.23}$, corresponding to a distance of 3.4 kpc (see Section \ref{redcl}), is indicated by a black vertical line. }
\label{lumfunc}
\end{figure}

Red clump (RC) stars can be used to obtain an independent estimate of the distance to the cluster. For GC01, a clumping of stars from  the red giant branch is seen around the apparent { dereddened} F153M magnitude of $\approx11.2$ (see Figure \ref{lumfunc}). In order to determine the distance corresponding to the magnitude at which the RC stellar distribution peaks, we use the { dereddened}  apparent luminosity function of GC01 in the F127M and F153M filters (in Figure \ref{lumfunc}, for brevity we only show the luminosity function for the F153M filter). These filters were chosen because they are the nearest to the Johnson $J$ and $H$ bands, where the red clump stars have well established absolute magnitudes (see \citeauthor{2016ARA&A..54...95G} \citeyear{2016ARA&A..54...95G} and references therein). Therefore, the conversion from  $J$ and $H$ magnitudes to F127M and F153M magnitudes are the least impacted by the assumptions about the RC stellar spectrum.

To convert between the {\sl HST} and Johnson filters we approximated the RC stellar spectrum using a PL model ($F_\lambda\propto \lambda^{\alpha}$) with a spectral index of $\alpha=-1.9$, consistent with the absolute magnitudes provided in \cite{2012MNRAS.419.1637L}. To ensure there are no broad absorption features in the RC stellar spectra at these wavelengths, we examined the corresponding Castelli \& Kurucz stellar atmosphere models \citep{2004astro.ph..5087C}. The stellar atmosphere models with solar and low metallicities are well approximated at these wavelengths by the assumed PL model. The absolute\footnote{All stars in the \cite{2012MNRAS.419.1637L} sample were nearby and showed no signs of extinction effects.} $J$ and $H$ magnitudes of \cite{2012MNRAS.419.1637L} converted to the F127M and F153M bands are --1.02 and --1.45, respectively\footnote{The magnitudes were converted using the power-law spectral model in pysynphot \citep{pysyn}.}.

In order to measure the distance to GC01, we found the apparent { dereddened} ($A_V=18$) magnitude of the RC stellar peak in the F127M and F153M luminosity functions. To ensure that the calculated distances were similar as a function of bin size, two different bin sizes (i.e., 0.1 and 0.2 magnitude per bin) were used. These two bin sizes were chosen to be small enough to capture the RC in the luminosity function, but large enough to have reasonably small statistical fluctuations per bin (see Figure \ref{lumfunc}). The apparent { dereddened} magnitudes corresponding to the RC are { 11.63--11.50} for the F127M filter, and { 11.23--11.11}, for the F153 filter. This corresponds to distances of 3.2--3.4 kpc. { Reddening} affects the distance derived from the RC position in the luminosity function. We repeated the above steps for $A_V=17$ and $A_V=19$ and found mean distances of 3.7 and 3.0 kpc, respectively.  

We also compared the luminosity functions of GC01 including ($r<42''$) and excluding ($18""<r<42''$) the cluster core and found an excess of RC sources in the core, suggesting possible mass segregation in GC01.  However, the large extinction and crowding can lead to biases. Future observations with {\sl JWST} can establish this much more confidently.

\subsection{Color-Magnitude Diagrams}
\label{cmdsec}

\begin{figure*}
\begin{center}
\includegraphics[scale=0.45,trim= 0 0 0 0]{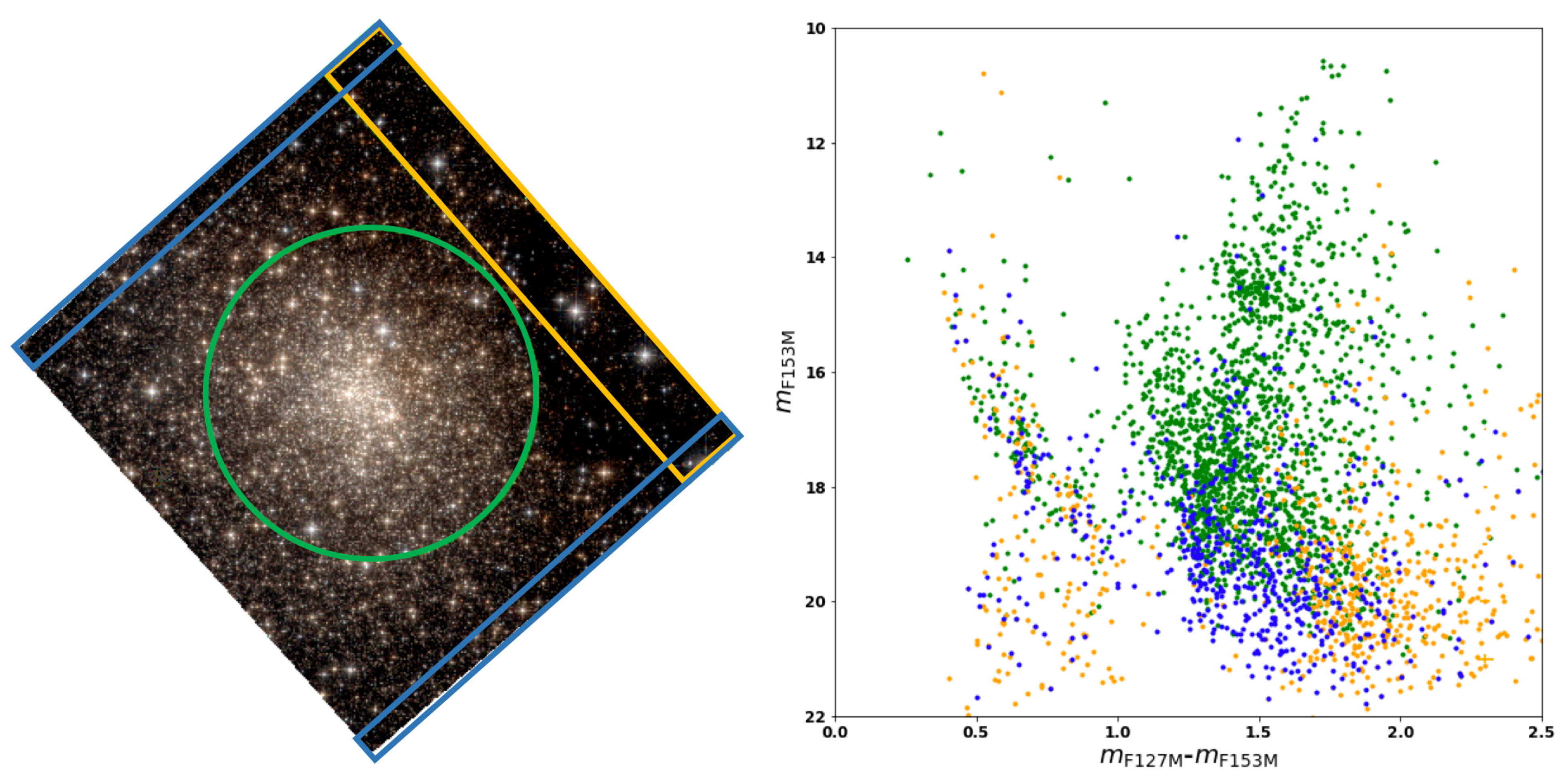}
\caption{{ {\sl Left:} False color WFC3/IR image (red: F153M, green: F139M, blue: F127M filters) image of GC01. North is up and East is to the left. The green circle shows the $r=42''$ ROI used for our analysis. The orange and blue boxes show the two different control field regions (Section \ref{cmdsec}). {\sl Right:} The observed CMD  ($m_{\rm F153M}$ vs. $m_{\rm F127M}$-$m_{\rm F153M}$)  for stars within the ROI (green) and for the two control fields (orange and blue). Cluster members tend to have $m_{\rm F127M}$-$m_{\rm F153M}$ colors of $\sim$1.2--1.7. The ROI has an area $\sim$2 times larger than each of the control fields.}}
\end{center}
\label{control}
\end{figure*}

In order to evaluate the contamination of the GC01 CMD by field stars we have defined a control field. Typically, this is done by imaging a flank field (see e.g., \citealt{2017A&A...602A..22A}), however, such an observation does not exist for GC01. Therefore, we chose a region, as far as possible from the cluster center, in the western part of the WFC3 images where the fraction of the cluster's stars appears to be minimal (orange rectangle in Figure \ref{control}).  Additional complications arise due to the presence of the dark dust lanes within the western half of the control field region.  This dust appears to be in front of the cluster, and therefore, { the contamination of the cluster's CMD by the faint background stars can be underestimated.  The stars from this control field appear to be either substantially ``bluer'' (foreground stars) or substantially ``redder''  (background stars, and possibly some of the cluster stars affected by larger extinction) than the stars from the cluster region  (see Figure \ref{control}).}

{ We have also evaluated the contamination by choosing two small slices to the north-east and south-west edges of the cluster as a control field (blue rectangles in Figure \ref{control}). In this case, we find substantially more control field stars overlapping with the cluster star region in the CMD plot. However, \cite{2016AJ....152..173D} have found that light from GC01 can be detected out to 100$''$ from the cluster center, suggesting that a large fraction of these ``contaminating stars'' are likely to be cluster stars (also see below). Therefore, either of the two choices of the control field introduces its own bias. However, in either case, the contamination of the cluster CMD at magnitudes brighter than  $m_{F153M}$=18 is minimal.} The crowding of sources in the dense cluster explains why the photometric catalog goes $\approx$2-3 magnitudes deeper in the less crowded control fields. The cluster stars lie primarily in the color space spanning $1.2\lesssim m_{\rm F127M}$-$m_{\rm F153M}\lesssim1.7$.  

\begin{figure*}
\begin{center}
\includegraphics[scale=0.47,trim = {10 0 0 0}]{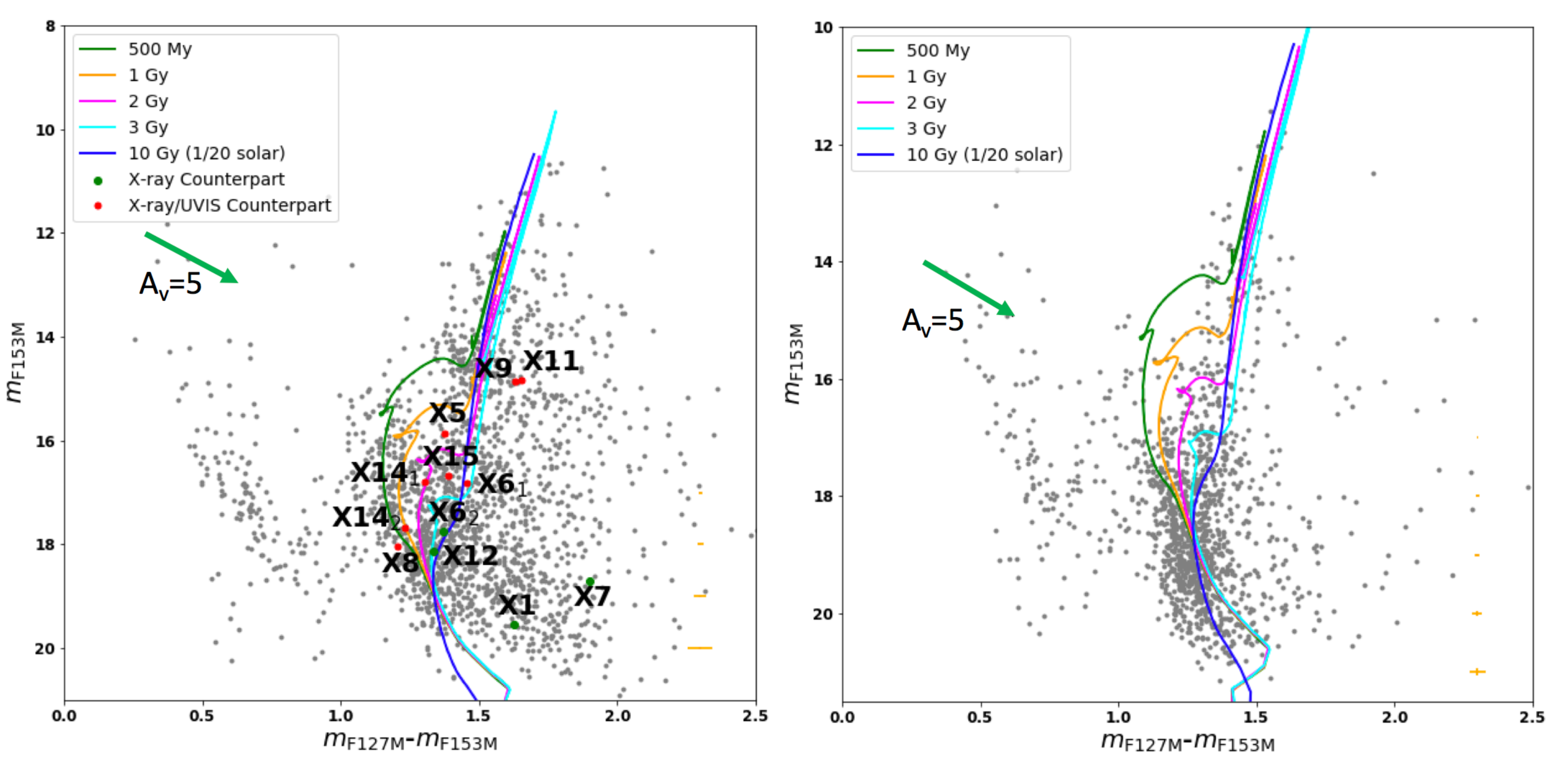}
\caption{{\sl Left:} Apparent WFC3/IR CMD  ($m_{\rm F153M}$ vs. $m_{\rm F127M}$-$m_{\rm F153M}$)  for stars within the $r=42''$ ROI with solar metallicity isochrones from 500 Myr to 3 Gyr and a 10 Gyr low-metallicity (1/20 solar) isochrone over-plotted. The isochrones are reddened with $A_V=18$, and placed at a distance of 3.3 kpc. The mean photometric color uncertainties are shown as orange error bars for several different magnitude The NIR counterparts to the X-ray sources are shown as colored (i.e., red or green) points and are numbered by a subscript if the X-ray source has multiple possible counterparts. The red points indicate that the source has both a WFC3/IR (SNR$>30$) and UVIS (SNR$>10$) counterpart, while the green points indicate that only a NIR counterpart is present. The green arrow shows a plausible differential { reddening} vector, $\Delta$A$_{V}=5$. {\sl Right:} Apparent WFC3/IR CMD  ($m_{\rm F153M}$ vs. $m_{\rm F127M}$-$m_{\rm F153M}$)  for stars within the dashed green box in Figure \ref{spitzer}. The isochrones are reddened with $A_V=17$ and placed at a distance of 3.3 kpc (see Sections \ref{colcol} and \ref{cmdsec}). { (Supplemental data for these figures are available in the online journal. In addition to the magnitudes shown, the data also include the magnitudes in the F814W and F139M filters, the magnitude uncertainties, and the SNR, crowding, and sharpness values for these sources in all filters.)}}
\end{center}
\label{cmds}
\end{figure*}

{ The left panel of Figure \ref{cmds} shows the $m_{\rm F153M}$ vs. $m_{\rm F127M}$-$m_{\rm F153M}$  apparent CMD together with 500 Myr, 1, 2, and 3 Gyr isochrones computed for solar metallicity and the 10 Gyr isochrone computed for 1/20 solar metallicity. The isochrones are reddened with an $A_V=18$ (see Section \ref{colcol}) and placed at $d=3.3$ kpc (see Section \ref{redcl}). The grey points are all  WFC3/IR sources that pass the photometric cuts described in Section \ref{hstphot}  and lie within the ROI. The representative photometric errors are shown as vertical and horizontal error bars near the right boundary  of each CMD for different magnitudes. Unfortunately, any rigorous fitting of the isochrones will be dominated by systematic uncertainties due to the large differential extinction, photometric crowding, and uncertain contamination from field stars. Therefore, it is difficult to set a constraining age limit on the cluster using the stars from the $r=42''$ ROI (see Section \ref{hstphot}).

In an attempt to obtain a more accurate  age constraint, we have carefully chosen a second cluster region (dashed green box in Figure \ref{spitzer})  offset from GC01's center which reduces the effects of crowding and differential reddening. The latter is accomplished by placing the region within the area of relatively uniform $^{12}$CO (J$=3\shortrightarrow2$) emission. We perform the same cuts on the photometric catalog as listed above (see Section \ref{hstphot}), leaving 1,354 NIR sources. We then followed the procedures described in Sections \ref{colcol} and \ref{redcl} to estimate the $A_V$ and distance in this new region (i.e., using the color-color diagram and RC stars), and find $A_V=16-17$ and $d=3.5-3.3$ kpc, respectively.

    The right panel of Figure \ref{cmds} shows the $m_{\rm F153M}$ vs. $m_{\rm F127M}$-$m_{\rm F153M}$ apparent CMD for WFC3/IR sources that lie within this alternative cluster region. The plotted isochrones are the same as in the left panel of Figure \ref{cmds} but are reddened with an $A_V=17$ and placed at $d=3.3$ kpc. The representative photometric errors are shown as vertical and horizontal error bars near the right boundary  of each CMD for different magnitudes. For solar metallicity, isochrones with ages $\gtrsim2$ Gyrs better match the CMD. Ages $\lesssim2$ Gyr are disfavored because these isochrones do not accurately capture the redder points between $m_{F153M}\approx15-17$, and there are a lack of sources extending up to the MSTOs at these ages. However, the 10 Gyr low-metallicity (1/20 solar metallicity) isochrone is also consistent with the data. Therefore, we cannot differentiate between an old ($\gtrsim10$ Gyr) low-metallicity globular cluster, and a younger ($\gtrsim2$ Gyr) solar-metallicity cluster with the current data. }

\begin{figure}
\begin{center}
\includegraphics[scale=0.42,trim = {35 0 0 0}]{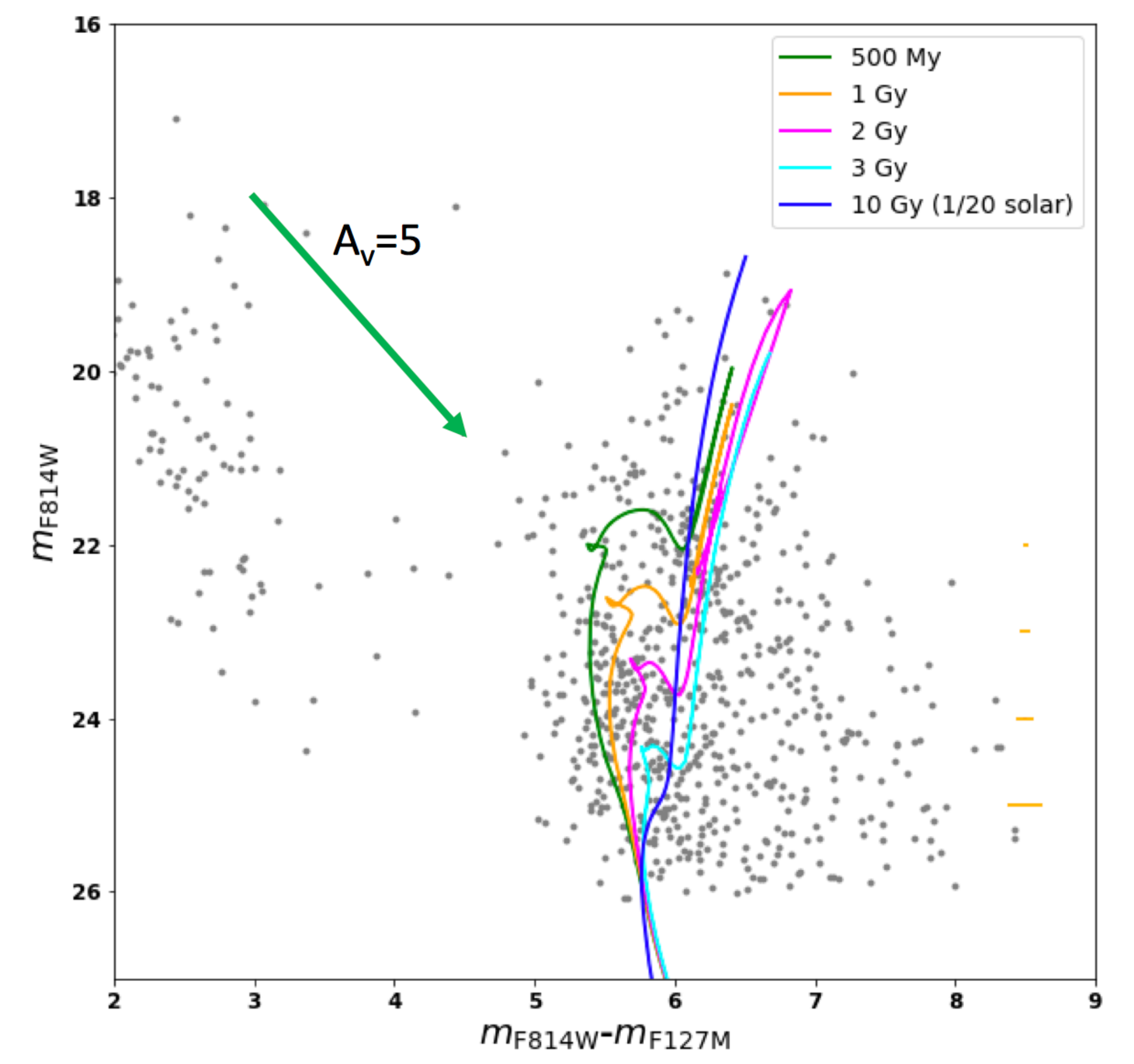}
\caption{Observed WFC3/UVIS and IR CMD  ($m_{\rm F814W}$ vs. $m_{\rm F814W}$-$m_{\rm F127M}$) for stars within $r=42''$ ROI with solar metallicity isochrones from 500 Myr to 3 Gyr, and a 10 Gyr low-metallicity (1/20 solar) isochrone over-plotted. The isochrones are reddened with $A_V=18$.  A distance of 3.3 kpc is assumed.  The mean photometric color uncertainties are shown as orange error bars for several different magnitudes.  The green arrow shows a plausible differential { reddening} vector, $\Delta$A$_{V}=5$. { (Supplemental data for this figure are available in the online journal. In addition to the magnitudes shown, the data also include the magnitudes in the F814W and F139M filters, the magnitude uncertainties, and the SNR, crowding, and sharpness values for these sources in all filters.)}}
\end{center}
\label{uviscmd}
\end{figure}

We have also constructed an optical-NIR CMD, { for stars in the $r=42''$ cluster ROI}, using the drizzled WFC3/UVIS image for source finding (see Figure \ref{uviscmd}). This approach reduces the number of confused (blended) NIR sources at the expense of having significantly less sources in the CMD. This CMD is in some sense the cleanest photometry we can produce { (although it still suffers from contamination by field stars)}, because of the higher spatial resolution and stricter cuts (the sources included in this CMD must pass both the NIR and UVIS cuts described in the last paragraph of Section \ref{hstphot}). Most of the counterparts to the X-ray sources do not pass the strict photometric cuts and hence do not appear in this CMD. The stretched appearance of the RC in this CMD emphasizes the effect of the differential { reddening. Although we show the same isochrones in Figure \ref{uviscmd} as we did in Figure \ref{cmds}, we cannot better constrain the age of GC01 from this CMD.}

White dwarf (WD) cooling curves could in principle be used to get an independent age estimate of the cluster (see e.g., \citealt{1998ApJ...504L..91R}). To estimate the observed magnitude of WDs in GC01 we used the WD sample provided in \cite{2006AJ....132.1221H}, observed in  2MASS's $J$, $H$, and $K$ bands. The $J$ and $H$ bands  overlap with the WFC3/IR  bands used in our observation (see Figure \ref{filter}). We took the hottest source in the above-mentioned WD sample (T$_{\rm eff}=92970$ K, d$=169$ pc, J$_{\rm mag}=15.303$, H$_{\rm mag}=15.604$) and placed it at the distance of  $3.3$ kpc with A$_J=5.1$ and A$_H=3.4$ \citep{1989ApJ...345..245C}. Converting from J and H bands to the F127M and F153M filters, this gives expected apparent magnitudes of $\sim27$ and $\sim25$, respectively\footnote{The conversion was done using a flat spectral model (i.e., ($F_\lambda\propto \lambda^{\alpha}$ with $\alpha=0$).}). Therefore, we do not expect to see any WDs belonging to the cluster in the IR/UVIS images, as the observations do not go deep enough.

\subsection{Discussion of Distance and Age }
\label{agediscuss}

{ The distance estimates using RC stars in the cluster's center implies a distance of $\approx$ 3.0--3.7 kpc depending on the absorption ($A_V=17-19$). This distance estimate is slightly more constrained { for the photometry from} the region outside of the cluster's core where the reddening is more uniform, giving $d=3.3-3.5$ kpc for $A_V=17-16$, respectively. This estimate is at the lower end of previous distance estimates of 3.1--5.2 kpc obtained from the ${}^{13}$CO feature emission and extinction map \citep{2005AJ....129..239K}, but is consistent with the 3.8$\pm0.7$ kpc distance obtained using RC stars \citep{2005A&A...442..195I}. However, these smaller distances still leave open the possibility that the cluster is  embedded in the ${}^{13}$CO cloud at 3.1 kpc.}

One of the most controversial properties of GC01 is its age. The isochrone matching to the cluster CMD suggests an age of $\gtrsim2$ Gyr, assuming a solar metallicity, $A_V=17$, and distance of 3.3 kpc. However, the 10 Gyr low metallicity isochrones are also consistent with the data. If GC01 does have a low-metallicity, it would then be more likely to be a a GC passing through the disk of the Galaxy, because low-metallicity GCs are typically found in the Galactic halo and not the Galactic disk (see e.g., \citealt{2013MNRAS.436..122L}). We find that ages $\lesssim2$ Gyr are inconsistent with the observed NIR CMD.  The WFC3/UVIS CMD also disfavors ages $<1$ Gyr.

 Recently, \cite{2016AJ....152..173D} have fitted {\sl SPITZER} IRAC CMDs and found  the age to be between 1 and 2.5 Gyr, compatible with our estimate for solar metallicity. Thus, it is possible that GC01 is a $\sim$2 Gyr old massive cluster born in the Galactic disk with an age similar to those of the Galactic clusters IC 4651, NGC 752, and M67 (\citealt{2002A&A...386..187M,1972MNRAS.157..147B,2015MNRAS.452.3394M}) but with a much larger mass. Due to its large mass, it could be an aged YMC, such as Westerlund 1, RSGC 03, or Arches \citep{2010ARA&A..48..431P}.  {  However, with the data at hand, we cannot rule out that GC01 is an old ($\sim$ 10 Gyr) low-metallicity globular cluster plunging into the disc (c.f., NGC 6544; \citealt{2017A&A...608A.140C})}. Future spectroscopic observations to determine the metallicity of GC01 would allow us to differentiate between these two scenarios. Further, the higher angular resolution  { and larger field of view} of {\sl JWST/NIRCam} can also help to better constrain the properties of the CMD of GC01.

\subsection{Loop-like Structure}
\label{loopstruc}

\begin{figure*}
\begin{center}
\includegraphics[scale=0.455]{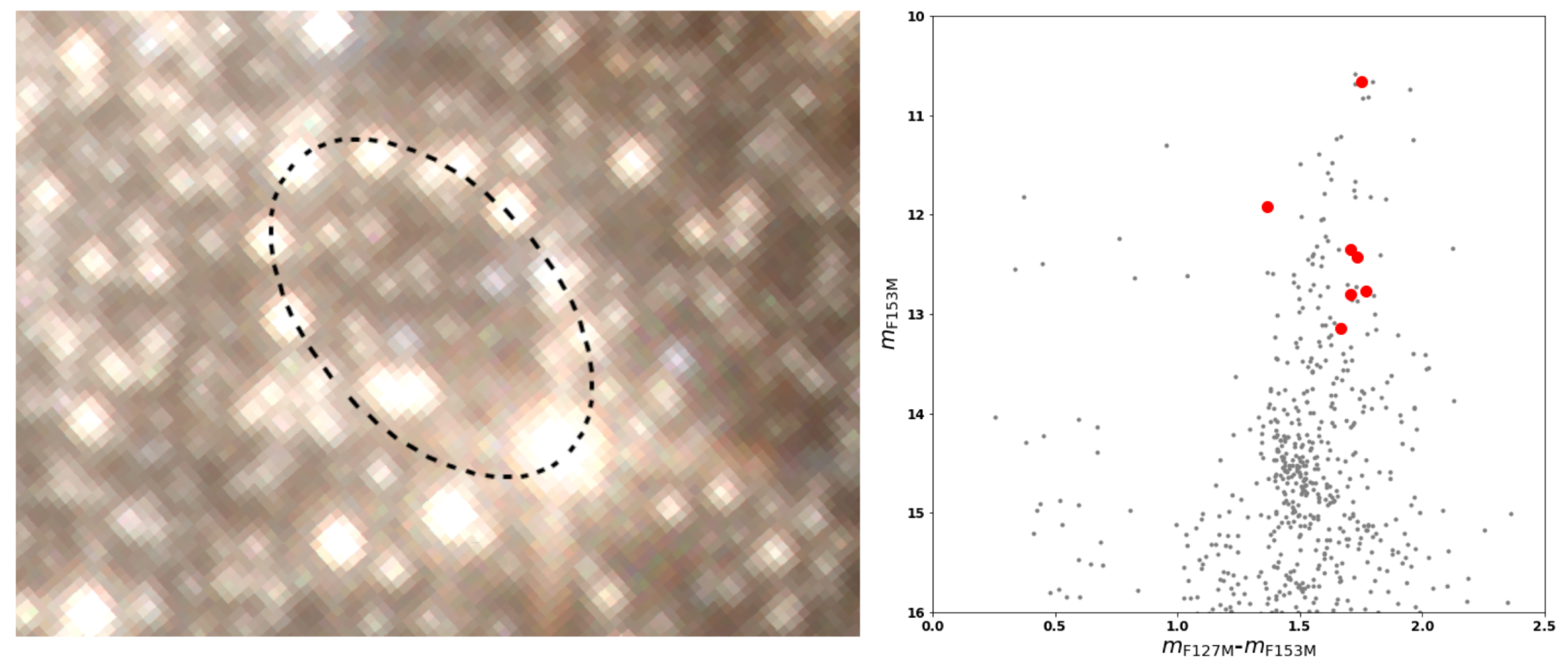}
\caption{Left: HST WFC3/IR false color image in F127M (blue), F139M (green), and F153M (red) filters featuring the loop-like structure discussed  in Section \ref{loopstruc}. Right:  $m_{\rm F153M}$ vs. $m_{\rm F127M}$-$m_{\rm F153M}$ CMDs showing the locations (with the red filled circles)  of the stars from  the loop-like structure. The stars are bright, and lie on the red side of red giant branch of the isochrones. }
\end{center}
\label{loop}
\end{figure*}

\cite{2005AJ....129..239K} reported a loop-like structure seen in the  {\sl Spitzer} IRAC images of GC01 from the {\sl GLIMPSE} survey. They rule out both dust shells ejected by stars and a supernova remnant due to the size of the feature and lack of radio emission, respectively. They suggest that the structure could either be an old nova shell or planetary nebula \citep{2005AJ....129..239K}. However, in the WFC3 images the structure is resolved into several stars of a similar brightness arranged in a peculiar loop-like pattern (see Figure \ref{loop}), which can be described as an ellipse with a semi-major axis $a=2\farcs7$ and semi-minor axis $b=1\farcs7$, corresponding to physical sizes of 0.04 pc and 0.03 pc, respectively (at a distance of 3.3 kpc). If the structure is a ring seen in projection onto the sky, the inclination angle would be $\simeq30^{\circ}$ { north through east}. 

Only seven out of $\sim13$  stars in the loop-like structure have  photometry that satisfies the quality criteria of the photometric catalog. These seven stars are plotted as red points on top of our CMD. They all lie towards the top of the CMD (see Figure \ref{loop}) and all but one  have a $m_{\rm F127M}$-$m_{\rm F153M}$ color of $\sim$1.7. Given the advanced age of the cluster it is difficult to imagine that the ring-like arrangement of the stars could be maintained since its formation. Most likely the structure is just an accidental arrangement in the projection onto the plane of the sky. Accurate 3D velocity measurements for these stars would provide further information. In addition to spectroscopic radial velocity measurements, future {\sl JWST} observations would allow one to measure tangential velocity components of the stars if the cluster is as close as 3.3 kpc.

\section{X-ray Sources}
\label{xray}

Here we discuss the most interesting sources individually, based on their X-ray properties, while the remaining X-ray sources are discussed in bulk. We compare them to the X-ray source populations found in both old open and globular clusters. { Ten} of the 15 X-ray sources labelled in the left panel of Figure \ref{ROI} have at least one NIR source located within their X-ray positional error circles, { while seven have at least one optical/NIR source positionally coincident\footnote{Including the exception of source X11, which has an optical/NIR source just outside of its $2\sigma$ X-ray positional uncertainty}.} { As mentioned above in Section \ref{chance}, we expect 9$\pm2$ or 4$\pm2$ X-ray sources to have NIR or NIR/optical counterparts due to chance coincidences, respectively. We do not insist that any one particular NIR/optical source is the true counterpart to an X-ray source. Below we rely primarily on the X-ray source properties to understand the nature of the X-ray sources. However, we do speculate on how the NIR sources coincident with the X-ray sources could be interpreted with respect to the X-ray source classification.}

  The unabsorbed 0.5--8 keV luminosities of the 15 X-ray sources span a range of $\sim10^{31}-10^{33}$ erg s$^{-1}$, with a majority having  $L_X=$1--5 $\times10^{31}$ erg s$^{-1}$ (at $d=3.3$ kpc). These  luminosities are higher than those of flaring MS stars\footnote{GC01 is  too old to have any pre-main sequence stars or star-forming activity.} found in old clusters (see e.g, \citeauthor{2008A&A...490..113G} \citeyear{2008A&A...490..113G}), implying that the detected sources can be a combination of ABs, MSPs (isolated or black-widow/redback type), CVs, and  quiescent LMXBs,  with the possible addition of more exotic source types,  such as aged magnetars or an intermediate mass black hole (IMBH)  accreting from the ISM (see Section \ref{imbh}).

  Given that { the age of GC01 is $\gtrsim2$ Gyr}, it could be that this cluster is an intermediate age cluster born in the disc. There are several open clusters with somewhat larger ages that have been observed with {\sl CXO}. M67 is an open cluster with an age of 4 Gyr  at a distance of 850 pc \citep{1999AJ....118.2894S}. This cluster has been observed by both {\sl CXO} and {\sl XMM-Newton} and hosts a number of ABs of different types, as well as a CV (\citeauthor{2004A&A...418..509V} \citeyear{2004A&A...418..509V}; \citeauthor{2015MNRAS.452.3394M} \citeyear{2015MNRAS.452.3394M}). NGC 6791 is an older open cluster (8 Gyr) that lies at a comparable distance of 4 kpc and has a mass of 5000-7000 M$_\odot$ (\citealt{2011ApJ...733L...1P}; \citealt{2013ApJ...770...98V}), { which is an order of magnitude smaller than the mass of GC01 ($8\times10^{4}$ M$\odot$; \citealt{2011MNRAS.411.1386D})}. NGC 6791 was observed with {\sl CXO} and a number of X-ray sources, including ABs, CVs and sub-subgiant { (SSG) binaries} were detected with unabsorbed 0.3--7 keV luminosities ranging between $(1-15)\times10^{30}$ erg s$^{-1}$ \citep{2013ApJ...770...98V}.  
 
Alternatively, if GC01 is a low-metallicity $\sim 10$ Gyr old cluster, then it could be a GC passing through the Galactic disk { (c.f., NGC 6544; \citealt{2017A&A...608A.140C})}. Many GCs have been observed with {\sl CXO}, such as 47 Tucanae \citep{2017MNRAS.472.3706B}, $\omega$ Centauri \citep{2009ApJ...697..224H}, Terzan 5 \citep{2003ApJ...590..809H} and M80 \citep{2003ApJ...598..516H}. In comparison to old open clusters, GCs host more MSPs (see e.g., \citealt{2017MNRAS.472.3706B}) and qLMXBs (\citealt{2003ApJ...598..501H}, \citealt{2012ApJ...745...57G}), while having fewer bright (i.e., $L_X>1\times10^{30}$ erg s$^{-1}$) CVs, ABs, and SSGs per unit mass \citep{2013ApJ...770...98V}.

In ABs, the X-rays come from interacting magnetically active stars. The magnetic activity of these stars is enhanced  due to a large rotation (i.e., periods on the order of 0.1-10 days \citealt{2004A&A...418..509V}) maintained by tidal interactions with the companion star \citep{2017ApJ...837..130V}. These sources typically account for a large fraction of the X-ray source population in old open clusters and often dominate their X-ray luminosity (see e.g., \citeauthor{2017ApJ...837..130V} \citeyear{2017ApJ...837..130V}, \citeauthor{2004A&A...418..509V} \citeyear{2004A&A...418..509V}).  
 The population of ABs in old open clusters have typical X-ray luminosities between 10$^{28}-$ a few $\times10^{31}$ erg cm$^{-2}$ s$^{-1}$ \citep{2015MNRAS.452.3394M, 2006ApJ...646L.143P} and soft spectra with median photon energies $E_{\rm med}\approx1.5$ keV \citep{2017ApJ...837..130V}. 
GCs also contain ABs (typically with $L_X<10^{31}$ erg s$^{-1}$), however, they have fewer per unit mass than open clusters \citep{2013ApJ...770...98V}.  This is because the AB population likely scales with the primordial binary fraction, which is reduced in GCs (see \citealt{2010AIPC.1314..135H} and references therein). GC01 may harbor many ABs, however, most of them will be too faint to be detected at our limiting luminosity.  
  
Very little is known about the population of MSPs in massive open clusters.  All known YMCs are too young to host MSPs spun-up by accretion.  On the other hand, most  known old open clusters are too scattered and probably do not have enough gravitational pull to retain isolated or even binary MSPs that acquire kicks from a supernova (SN) explosion \citep{2011ApJ...743..102G}.
 
If GC01 is a GC then it could host MSPs, as GCs are known to be rich with them (e.g., \citealt{2006ApJ...646.1104B,2017ApJ...845..148P}). Typically, MSPs exhibit a thermal ($kT\approx$0.1--0.2 keV) X-ray spectrum, but several also have a hard power-law component ($\Gamma=1-1.5$) and may show orbital variability on times-scales of a few to tens of hours (see e.g., \citealt{ 2006ApJ...646.1104B,2011ApJ...730...81B,2014ApJ...795...72L}). However, due to the large absorption towards GC01, we would not detect the lower luminosity ($L_{0.5-8}<1.5\times10^{32}$ erg s$^{-1}$) thermal MSPs in the existing data.

Quiescent LMXBs  are another type of relatively faint X-ray sources expected in evolved stellar environments. \cite{2003ApJ...598..501H} studied $\sim20$ qLMXB candidates detected in nine different GCs and found an empirical lower limit  of 10$^{32}$ erg s$^{-1}$ on their X-ray luminosity. In X-rays, qLMXBs with NSs appear as fairly soft sources, due to the thermal emission from the NS (e.g., Table 2 in \citealt{2003ApJ...598..501H} and Figure 1 in \citealt{2018arXiv180107107S}). There have also been a number of BH qLMXB candidates with $L_{X}>10^{32}$ erg s$^{-1}$ recently discovered in GCs (see e.g., \citealt{2013ApJ...777...69C,2015MNRAS.453.3918M,2017MNRAS.467.2199B}) based on their X-ray to radio luminosity ratios. While in quiescence, BH LMXBs often exhibit hard spectra ($\Gamma=1-2$) making them similar to CVs, but a CV nature can be ruled out if a radio counterpart is detected \citep{2017MNRAS.467.2199B}. At fainter luminosities, \citep{2014MNRAS.444..902A} reported a confirmed BH qLMXB with $L_X= 8\times10^{29}-1.3\times10^{31}$ erg cm$^{-2}$ s$^{-1}$ and a PL photon index $\Gamma=2.1$. Therefore, qLMXBs with BHs may be detectable in the existing {\sl CXO} observation of GC01, but are difficult to identify without deep X-ray and radio observations.

 To better understand the nature of GC01's X-ray sources, we carried out spectral fitting and evaluated the X-ray properties of the brightest sources.
Below we discuss the most interesting sources and the different classes they could belong to. Unfortunately, the {\sl HST} and X-ray observations of GC01 are too short to search for periodicity associated with binary motion and the X-ray sources have too few counts to search for a spin period, or in many cases, fit the spectra. The orbital period of the system, along with the X-ray luminosity and spectral slope, can help to determine the nature of the source (see e.g., \citeauthor{2004A&A...418..509V} \citeyear{2004A&A...418..509V}) and deeper X-ray, NIR, and radio observations of GC01 should be undertaken. 

\subsection{Source X1}
This source has 61 net counts and an observed X-ray flux of $3.6\times10^{-14}$ erg s$^{-1}$ cm$^{-2}$. The spectrum is hard and can be described by an absorbed PL with $\Gamma=0.9\pm0.3$ at a fixed $N_{\rm H}=4\times10^{22}$\,cm$^{-2}$. The  unabsorbed luminosity of X1 is  $L_{\rm 0.5-8}=9.1\times10^{31}$ erg s$^{-1}$ at $d=3.3$ kpc.  The X-ray luminosity and spectral hardness of X1 are consistent with a CV or qLMXB nature of the source. The best-fit $\Gamma$ is smaller than  typical MSP or qLMXB values, but the uncertainties are large. However, for an MSP scenario, the non-thermal luminosity is too large. X1 has an X-ray color, $X_{\rm color}=-0.1$, and luminosity, { $L_{\rm 0.5-6}\approx7\times10^{31}$ erg s$^{-1}$}, which are  representative of CVs according to  Figure 1 of  \cite{2006ApJ...646L.143P}, with ABs being less probable because they typically have lower luminosities and softer X-ray spectra.

  X1 has only one NIR { source} within the 2$\sigma$ X-ray position error circle with no UVIS counterpart (see Fig. \ref{cutout}). { If the NIR source is the true counterpart to the X-ray source}, the X-ray to NIR flux ratio, { $f_{X}/f_{\rm F127M}\approx0.08$}, is consistent with a CV interpretation (see Figure \ref{irxrat}). Unfortunately, X1 has a high crowding value and could be a mix of two or more faint sources, which would decrease the X-ray to NIR flux ratio. 

\subsection{Source X2}
{ Source X2 is the brightest source detected in GC01 with an observed flux of $1.2\times10^{-14}$ erg s$^{-1}$ cm$^{-2}$ { in the 0.5-8 keV band. It} was classified as a qLMXB by \cite{2007arXiv0708.3365P}}.  Most of the flux is in the soft band, as evidenced by both its large photon index $\Gamma=5.6\pm0.6$ and X-ray color $X_{\rm color}=2.1$. The unabsorbed X-ray luminosity is $L_{\rm 0.5-8}=3.4\times10^{33}$ erg s$^{-1}$ at a distance of 3.3 kpc and $N_{\rm H}=4\times10^{22}$\,cm$^{-2}$. In GCs, where qLMXBs are most commonly found \cite{2003ApJ...598..501H}, sources with luminosities $>10^{32}$ erg s$^{-1}$ and soft X-ray spectra are typically qLMXBs \citep{2006ApJ...651.1098H}. { For comparison to other qLMXBs, we fit source X2's X-ray spectrum with a hydrogen-atmosphere neutron star model (with fixed N$_{\rm H}=4\times10^{22}$ cm$^{-2}$, M$=1.4M_{\odot}$, $R_{\rm NS}=10$ km, and $d=3.3$ kpc). The best-fit temperature is $T_{eff}=0.13\pm0.03$ keV with an unabsorbed X-ray luminosity $L_X=4\times10^{32}$ erg  s$^{-1}$. The source's X-ray color, calculated using the neutron star atmosphere model, is $X_{\rm color}=1.3$, which is consistent with other known qLMXBs (see Figure 1 in \citeauthor{2006ApJ...646L.143P}  \citeyear{2006ApJ...646L.143P}).}
 
{ In open clusters, qLMXBs are quite rare, but one has been found in NGC 6819. This cluster has an age $\sim2$ Gyr \citep{2012ApJ...745...57G}, mass $\sim2600$ M$_{\odot}$, and half-light radius 3\farcm3 \citep{2001AJ....122..266K}, corresponding to a diameter of 4.4 pc at its distance of 2.3 kpc \citep{2011ApJ...729L..10B}. Comparatively, GC01 is more dense than NGC 6819, having a diameter $\sim1$ pc (at a 3.3 kpc distance, see Section \ref{redcl}) and mass $\sim8\times10^{4}$ M$_{\odot}$ \citep{2011MNRAS.411.1386D}. Therefore, it is plausible that GC01 could host a dynamically formed qLMXB regardless of whether it is a globular or open cluster.} 

X2 has no NIR/optical counterpart within its X-ray positional error uncertainty. It is common for qLMXBs in globular clusters to have very faint optical counterparts (see \citeauthor{2003ApJ...598..501H} \citeyear{2003ApJ...598..501H}). { Deep IR observations with {\sl JWST} can help to reveal the NS's companion and place constraints on its mass.}

\subsection{Sources X5}

This source has an X-ray color $X_{\rm color}=-0.3$, unabsorbed X-ray luminosity is $L_{\rm 0.5-6}=3.4\times10^{31}$ erg s$^{-1}$ at a distance of 3.3 kpc, suggesting a CV nature of the source (however, see Section \ref{imbh}). { Source X5 appears to be coincident with a single NIR source, which is resolved by UVIS into two sources, one of which falls close to the center of X-ray error circle (see Figure \ref{cutout}). Additionally, the NIR magnitude of the unresolved source has a large crowding value, suggesting that the photometry is not accurately recovered. Deeper images with a higher angular resolution detector, such as {\sl JWST's} NIRcam, are necessary to accurately recover the NIR properties of these potential counterparts.}


\subsection{Source X8}
Source X8 has an unabsorbed X-ray luminosity $L_{\rm 0.5-6}=1.3\times10^{31}$ erg s$^{-1}$, and X-ray color $X_{\rm color}=-0.1$, { suggesting a CV or AB nature.} This source is also one of the furthest from the cluster center. { The coincident NIR ($m_{F127M}$=19.2) and optical source ($m_{F814W}$=24.0) of X8  is blue ($m_{F814W}-m_{F127M}\approx$4.8) compared to the main locus of cluster stars in the CMD. This suggests that the NIR/optical source may be unrelated to the X-ray source. However, if the optical/NIR sources is the true counterpart, then given its blue color, and distance from the cluster center, X8 could be a foreground star with the X-rays being produced in the stellar corona}.


\subsection{Source X9}
\label{src9}
{ Source X9's soft X-ray color ($X_{\rm color}=$0.4) and luminosity ($L_X=3\times10^{31}$ erg cm$^{-2}$ s$^{1}$), suggests that it could be an AB. Additionally, this source is coincident with a NIR source, which is bright ($m_{F127M}$=16.5) with well measured photometry. We note that this NIR source lies on the redder side of the red giant branch, where there is very little contamination from field stars. The large differential reddening can ``extend" the red giant branch to this region of the CMD. Alternatively, this region of the CMD can also be populated with SSG or red straggler (RS) type ABs. SSGs and RSs lie on the redder side of the horizontal branch with SSGs being brighter and RSs being fainter than sub-giant branch stars \citep{2017ApJ...840...66G}. Typically, SSG and RS type { ABs} have X-ray luminosities of  $10^{30}-10^{31}$ erg s$^{-1}$ due to coronal magnetic activity \citep{2017ApJ...840...66G}, consistent with source X9's X-ray properties. These types of sources have been found in both open and globular clusters \citep{2017ApJ...840...66G}.} 



\subsection{Source X11}
\begin{figure}
\includegraphics[scale=0.38]{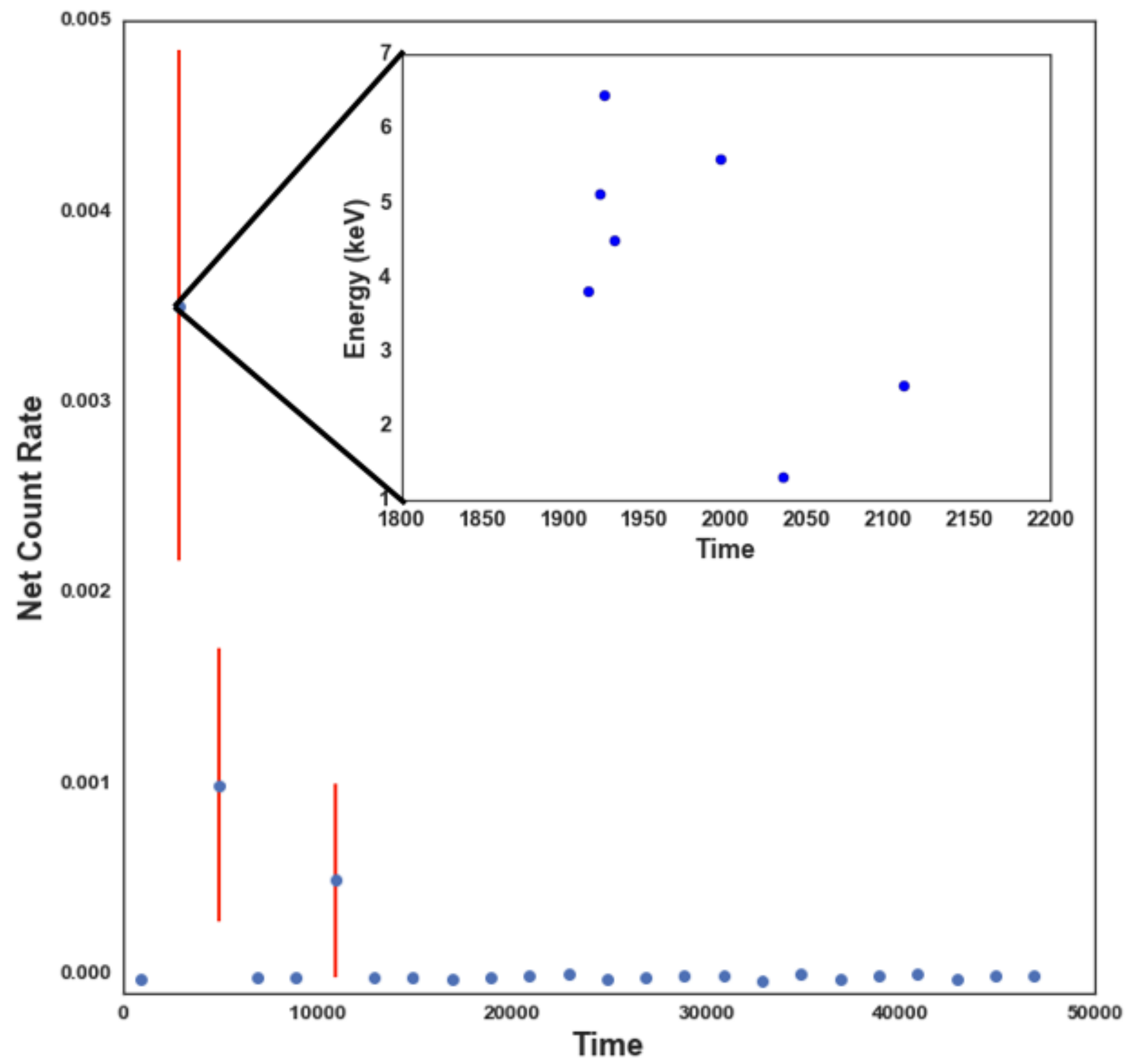}
\caption{X-ray light curve with 2 ks bins for the only variable X-ray source (X11). The inset shows the energies of the photons arriving during the 200 s flare.}
\label{src10xraylc}
\end{figure}

{ X11, reported as variable by \cite{2007arXiv0708.3365P}, was only detected with ACIS  during a 200-s flare and subsequent decay over the next 8 ks (see Figure \ref{src10xraylc}). At its peak, the flare reached a luminosity of $L_{\rm 0.5-8}=2.1\times10^{33}$  erg s$^{-1}$ at a distance of 3.3 kpc. This suggests that X11 could be either an AB or CV, as  both have exhibited X-ray flares reaching these luminosities (see e.g., sources W47 and W51 in 47 Tuc \citealt{2005ApJ...625..796H}). The fact that the source shows no persistent X-ray emission down to luminosities $L_X\sim10^{31}$ erg cm$^{-2}$ s$^{-1}$, suggests that it is more likely to be an AB type system but several X-ray faint ($L_X\approx10^{30}$) CVs have recently been identified in 47 Tuc \citep{2018MNRAS.475.4841R}.}

{ There is no NIR/optical counterpart within the 2$\sigma$ X-ray positional error circle, however, there is a NIR source right on the very edge of the error circle (See Figure \ref{cutout}). This source is bright ($m_{F127M}$=16.5), has reliable photometry, and lies very close to X9's potential counterpart in the NIR CMD (see Figure \ref{cmds}, left panel). Therefore, this source could be similar to X9 in its nature (SSG/RS type AB), assuming both NIR counterparts are real. However, SSG/RS are not known to show flares with such high luminosities. On the other hand, if the NIR source is not a true counterpart to X11, then the X-ray source could still be an LMXB, AB, or CV with episodic accretion and a faint non-detected companion.}


\subsection{Remaining Sources}
\label{remainder}

The remaining eight X-ray sources have unabsorbed luminosities of $\sim0.8-5\times10^{31}$ erg s$^{-1}$  at a distance of 3.3 kpc and X-ray colors between -0.8 and 0.4. Sources with $X_{\rm color}<0.0$ in this luminosity range tend to be CVs, while those with $X_{\rm color}>0.0$  can be a mix of ABs and binary MSPs \citep{2006ApJ...646L.143P}. 

\begin{figure}
\includegraphics[scale=0.36, trim={30 0 0 0}]{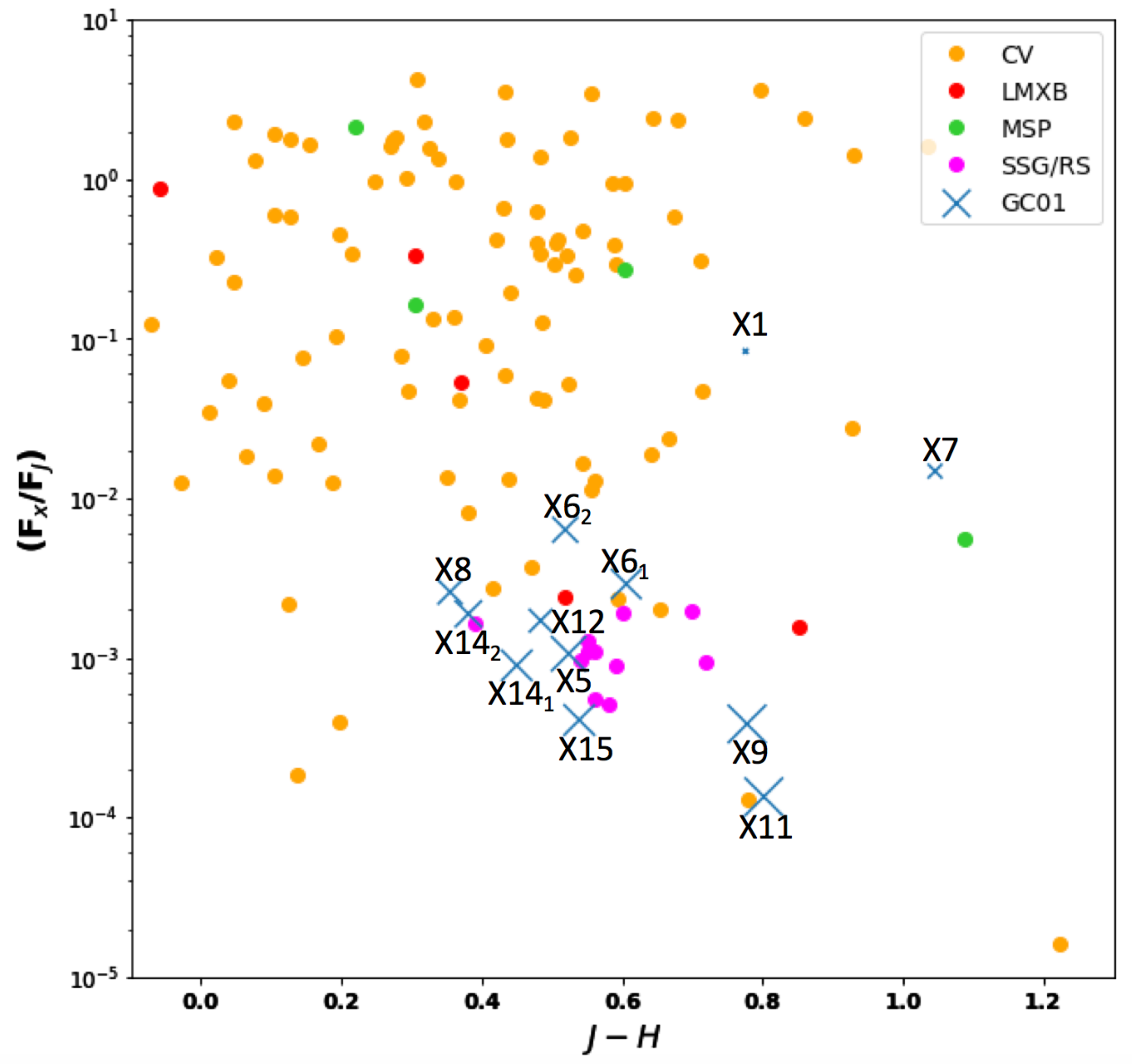}
\caption{ $\rm F_{\rm X}/F_{J}$ versus $J-H$ color diagram. The potential NIR counterparts to the X-ray sources were { dereddened} with an A$_V=18$ and are plotted { as blue crosses. The size of the crosses are proportional to the the brightness of the potential NIR counterpart to the X-ray source (i.e., brighter potential counterparts are shown as larger crosses).} The CVs (orange), LMXBs (red), MSPs (green), and SSG/RS (magenta) have been taken from various catalogs and the literature and are all detected in X-rays and NIR (see Section \ref{remainder}). It should be noted that the colors can be very uncertain for NIR sources with large crowding values (see Table \ref{counterparts}).}
\label{irxrat}
\end{figure}

In an attempt to classify these sources we have plotted the distance independent { dereddened} $J-H$ color versus the X-ray (0.5-7 keV) to $J$ band flux\footnote{Both fluxes were corrected for absorption.} ratio for the GC01 X-ray sources and candidate NIR counterparts (see Figure \ref{irxrat}). { We also plotted a number of CVs and LMXBs from the Cataclysmic Variables catalog \citep{2001PASP..113..764D} and Low-Mass X-ray binary catalog \citep{2007A&A...469..807L}, respectively, that have been detected by {\sl XMM-Newton} \citep{2001A&A...365L...1J} and have counterparts in 2MASS (see the Appendix of \citealt{2016ApJ...816...52H} for more details).} The X-ray fluxes and the $J$ and $H$ band photometry were retrieved from the 3XMM-DR6 catalog \citep{2016A&A...590A...1R} and 2MASS catalog of point sources \citep{2003tmc..book.....C}, respectively. Additionally, a sample of one globular and 10 open cluster SSG/RS type stars, which are detected at NIR ($J$ and $H$ bands) and X-ray wavelengths, was taken from Table 4 in \cite{2017ApJ...840...66G}. { We have also used four binary MSPs}\footnote{We could not use the large sample of MSPs detected in GCs because they do not appear in 2MASS and do not have NIR magnitudes.}, that have been detected by 2MASS and in X-rays. { We converted the X-ray fluxes of the MSPs into the 0.5-7 keV energy range using the spectral models in the respective papers of each source (\citealt{2011ApJ...742...97B, 2006ApJ...638..951Z, 2014ApJ...789...40B, 2002ApJ...569..894Z}).} The F127M and F153M magnitudes of GC01's NIR sources were converted to the $J$ and $H$ bands, respectively. The GC01 sources primarily fall near CVs and SSG/RS type ABs in this plot. 

\begin{figure}
\includegraphics[scale=0.3, trim={20 0 0 0}]{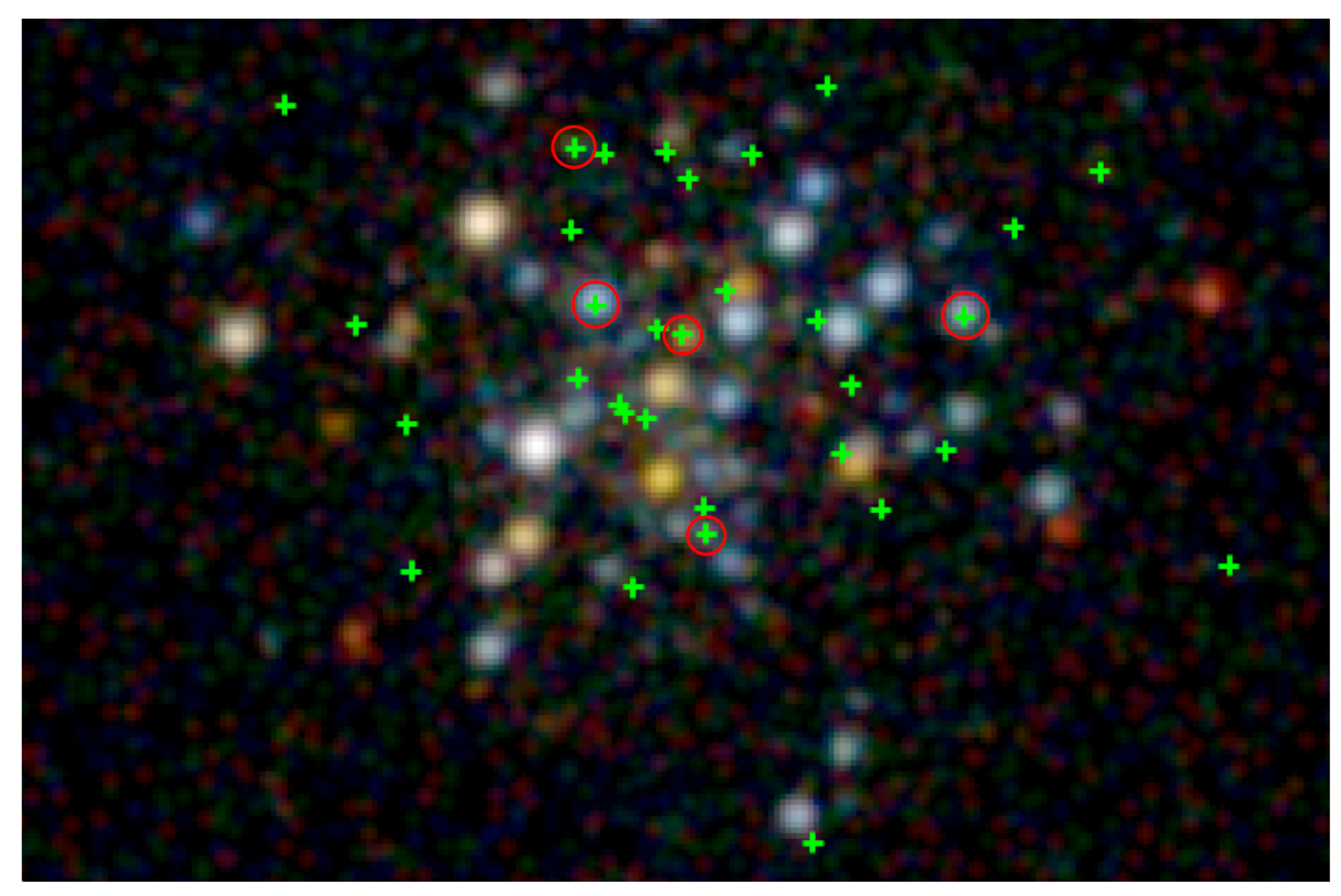}
\caption{Merged 530 ks \emph{CXO} false color image ($3-8$ keV - blue, $1.5-3$ keV -- green, and $0.5-1.5$ keV -- red) image of globular cluster Terzan 5. North is up, East is to the left. The data are smoothed (with a Gaussian kernel with a radius of 2\farcs0). The green crosses show the radio positions of the MSPs reported in \cite{2017ApJ...845..148P}, while the red circles highlight radio MSPs that are coincident with  X-ray point sources. Most of the X-ray sources seen in this image are likely to be CVs and qLMXBs (see e.g., \citealt{2006ApJ...651.1098H}).}
\label{T5MSPs}
\end{figure}

There are large variations in the fraction of MSPs detected in X-rays among GCs. For instance, 47 Tuc hosts 25 radio MSPs, of which 22 ($\sim90\%$) are detected in X-rays (see combined 540 ks {\sl CXO} ACIS image in \citealt{2017MNRAS.472.3706B}). On the other hand, for Terzan 5 (combined 530 ks {\sl CXO} ACIS exposure, see Figure \ref{T5MSPs}) there are 36 known radio pulsars \citep{2017ApJ...845..148P} of which only about 5 (or 14\%) appear to be detected in X-rays  (see Figure \ref{T5MSPs}). The most likely reason for such a drastic difference is the much larger absorption ($N_{\rm H}=1\times10^{22}$ cm$^{-2}$ and $A_V=6.7$) towards Terzan 5, which is strongly affecting the detectability of the MSPs, most of which have thermally dominated spectra (see e.g., \citealt{ 2006ApJ...646.1104B}).  Given that  GC01  is even more obscured (by a factor of 4) than Terzan 5, it is not surprising that in the existing short {\sl CXO} exposure we would see few or even no MSPs. { Additionally, no MSPs have been detected at radio wavelengths in GC01 so far (S. Ransom, private comm.).} Better quality X-ray and radio data for GC01 will allow us to directly compare the population properties of X-ray sources in Terzan 5 and GC01.

\subsection{Intermediate mass BH limits}
\label{imbh}

There have been theoretical predictions that intermediate-mass BHs (IMBH) could exist in the centers of GCs (see \citealt{2004cbhg.symp...37V} for a review). Regardless of whether GC01 is an old open or globular cluster, it still has a very large core density and hence, may host an IMBH. If massive main sequence stars in GC01 undergo fast mass segregation, they can develop a Spitzer instability and N-body simulations show that this can lead to a core collapse time $\sim0.1$ of the initial half-mass relaxation time \citep{2004ApJ...604..632G}. If the massive stars' lifetimes are shorter than the core collapse time, the stars can collide and/or merge leading to a runaway collision process and, possibly, the formation of an IMBH \citep{2002ApJ...576..899P,2006MNRAS.368..121F}. \cite{2002ApJ...576..899P} find that clusters with present day half-mass relaxation times $<10^8$ yrs and ages exceeding 25 Myr could contain an IMBH. N-body simulations predict that such an IMBH can have a mass of $\sim0.1\%$ of the birth mass of the cluster { \citep{2002ApJ...576..899P,2004ApJ...604..632G}}. \cite{2011MNRAS.411.1386D} estimate that the relaxation time of GC01 is $\sim10^{8}$ yrs. Further, the current mass of GC01 could be as large as $\sim80,000$ M$_{\odot}$, implying a birth mass up to an order of magnitude larger than its current mass \citep{2011MNRAS.411.1386D}. This suggests that a central IMBH could be as massive as $\sim800$ M$_{\odot}$.

If there is an IMBH in GC01 it should accrete from the ISM in the cluster and may produce detectable emission. Emission from isolated stellar mass and IMBHs accreting  at very low rates is poorly understood. X-ray observations of BH qLMXBs accreting at low rates show  X-ray spectra which can be described as a PL with photon indices $\Gamma\approx2$ \citep{2014MNRAS.444..902A,2016MNRAS.456.2707P}. Starved supermassive BHs (such as Srg A$^{\ast}$) seem to radiate most of their energy at sub-mm wavelengths (via synchrotron emission) with a possible second less energetic peak in X-rays due to inverse Compton (see \citealt{2017MNRAS.468.2447P} and references therein). 

Since the putative IMBH would be in the cluster center, we consider X-ray sources X9 and X14, (lying 1-2$''$ away) as potential IMBH candidates. Source X9 is a likely SSG/RS type system, while source X14 has two potential NIR counterparts with photometry affected by the crowding. Source X5 is also only $\sim4''$ away from the cluster center, but this source has a rather hard spectrum, making it a less likely IMBH candidate assuming the IMBH X-ray spectra are similar to those of low-$L_X$ qLMXBs. All three sources have $F_X/F_J=10^{-4}$--$10^{-3}$. Comparatively, Sgr A$^{\ast}$ and the quiescent BH binary {\sl Swift} J1357.2-0933 have X-ray to NIR flux ratios $F_X/F_J\approx10^{-3}$ and $10^{-1}$, respectively. Their X-ray luminosities are a tiny fraction of the Eddington luminosity, $\sim10^{-9}L_{\rm Edd}$ \citep{2016A&A...589A.116M, 2016MNRAS.456.2707P}. An 800 M$_{\odot}$ quiescent BH in GC01 with an X-ray luminosity of 10$^{-9}L_{\rm Edd}\approx10^{32}$ erg cm$^{-2}$ s$^{-1}$ would have an absorbed X-ray flux $F_{x}=3\times10^{-14}$ erg cm$^{-2}$ s$^{-1}$ (for $\Gamma=2$, N$_{\rm H}=4\times10^{22}$ cm$^{-2}$), which would be detectable in the {\sl CXO} observation. { Of course, the regimes of accretion in these two low-efficiency systems can still be quite different from that of a putative IMBH in GC01.} Deeper X-ray and NIR/IR observations (including NIR spectroscopy) are needed to probe the nature of X14, X9, and X5 to see if one of them could be an IMBH candidate in GC01.

\section{Outlook}
\label{outlook}

There are two major obstacles to performing a more informative study of GC01. The first is the strong crowding in the cluster's core. This can be overcome with JWST's superior angular resolution, which will produce images with a quality comparable to that of the WFC3/UVIS F814W images. {\sl JWST} will resolve sources that are too close together (such as the NIR counterpart to X5), provide accurate photometry, and be less affected by the large { reddening}. The second obstacle is the size of the X-ray error circles { and faintness of the X-ray sources}. Deeper X-ray observations will allow us to detect more X-ray sources that can be cross-matched directly to {\sl HST} and {\sl JWST} sources and can then be used to improve the relative astrometry. This will reduce { both the statistical and systematic} X-ray positional uncertainties allowing for more credible counterpart matching. { Additionally,} deeper X-ray observations would also better characterize the source spectra and allow for variability studies. { These goals will be achieved in the forthcoming deep CXO observation of GC01.} It is important to understand the nature of the NIR and X-ray sources in GC01, as it may be only one of a few massive intermediate age clusters observable in the Galaxy and can provide hints to both its formation and evolution. Finally, it is critical to perform sensitive low-frequency radio observations to look for MSPs that are expected to be numerous if GC01 is an old GC.

\section{Summary}
\label{summ}

Using both {\sl HST} and {\sl CXO} we were able to probe the parameters of GC01 and its X-ray source population. { In the cluster center}, we have detected 1,964 sources in the WFC3/IR F127M, F139M, and F153M  images and { 777} sources in the WFC3/UVIS F814W image. A color-color diagram suggests a variable extinction $A_V=18\pm4$. { We have also selected a region with less differential reddening ($A_V=16-17$) and crowding just outside of the cluster's core, and have analyzed 1,354 NIR sources in this region.} The peak of the red clump star distribution { in this alternative cluster region} was used to estimate the distance to the cluster, giving { $d=$3.5--3.3 kpc for $A_V=16-17$, respectively}. The CMDs and stellar isochrones corrected for this extinction and placed at a distance of 3.3 kpc imply an age of { $\gtrsim2$} Gyrs { up to $\sim$10 Gyrs for a lower metallicity.} Therefore, the {\sl HST} photometry by itself does not confidently discriminate between an intermediate age massive galactic cluster or an old globular cluster plunging into the disk. The lack of any radio MSPs could suggest that GC01 is not an old GC (such as Terzan 5 or 47 Tuc), although there remains a possibility that their detections are hampered by the large absorption towards GC01. 

We have analyzed the 15 brightest X-ray sources located within the central part of GC01. Nine of the X-ray sources have at least one coincident NIR source seen by {\sl HST} and three have multiple coincident NIR sources. { Additionally, seven of the X-ray sources are coincident with at least one optical/NIR source. This suggests that several of the optical/NIR sources are true counterparts of the X-ray sources. However, we cannot confidently determine which sources in particular are true counterparts to the X-ray sources.} Source X1 is a likely CV candidate given its X-ray colors and luminosity. One of the sources (X2) is a likely a qLMXB as it has a soft X-ray spectrum ($\Gamma=5.6$) and relatively large { unabsorbed} X-ray  luminosity ($L_{\rm 0.5-8}=3.4\times10^{33}$ erg s$^{-1}$). X9 is likely an AB type system with a { potential} NIR counterpart that has well measured photometry that suggests it { is a red giant}.

X11 showed an X-ray flare that lasted about 200 s and reached a peak X-ray luminosity of $L_{0.5-8}=2.1\times10^{33}$ erg s$^{-1}$ and then decayed back to below the detection limit over 8 ks. { This source is likely to be either an AB type system or CV}. X11 does not have a NIR/optical counterpart inside of its X-ray positional error circle. However, there is one NIR/optical counterpart, with well measured photometry that lies in the { red giant} region of the CMD, just on the edge of the X-ray positional error circle and may still be associated with X11.

\noindent{ Acknowledgements:} Support for Program number HST-GO-14183.003-A  was provided by NASA through a grant from the Space Telescope Science Institute, which is operated by the Association of Universities for Research in Astronomy, Incorporated, under NASA contract NAS5-26555. We are very grateful to Leisa Townsley and Patrick Broos for their help with running {\sl ACIS Extract}, as well as for insightful discussions on X-ray image analysis. { We would like to thank the anonymous referee for their helpful and constructive comments, which improved the paper.}


\end{document}